\documentclass[11pt]{article}

\usepackage[T1]{fontenc}
\usepackage[cp1251]{inputenc}
\usepackage{textcomp}
\usepackage[centertags]{amsmath}
\usepackage[mediummath]{nccmath}
\usepackage{amsfonts}
\usepackage{amssymb}
\usepackage[mathcal]{euscript}
\usepackage{braket}
\usepackage{tikz-feynman} 
\tikzfeynmanset{compat=1.0.0} 
\usepackage{simplewick} 
\usepackage[pdftex]{hyperref}
\usepackage{graphicx}
\usepackage[numbers,sort&compress]{natbib}

\usepackage{paperinitial}

\paperinitialization{15mm}{15mm}{15mm}{15mm}{2pt}{10pt}

\DeclareMathOperator{\re}{Re}
\DeclareMathOperator{\im}{Im}

\DeclareMathOperator{\Texp}{Texp}
\DeclareMathOperator{\Sp}{Sp}

\newcommand{\lan}{\langle}
\newcommand{\ran}{\rangle}
\newcommand{\bs}{\boldsymbol}

\newcommand{\e}{\varepsilon}
\newcommand{\vf}{\varphi}

\newcommand{\s}{\sigma}
\newcommand{\Si}{\Sigma}
\newcommand{\al}{\alpha}
\newcommand{\be}{\beta}
\newcommand{\ga}{\gamma}
\newcommand{\Ga}{\Gamma}
\newcommand{\de}{\delta}
\newcommand{\De}{\Delta}

\newcommand{\la}{\lambda}

\newcommand{\spx}{\mathbf{x}}

\newcommand{\spp}{\mathbf{p}}
\newcommand{\spq}{\mathbf{q}}
\newcommand{\spg}{\mathbf{g}}
\newcommand{\spk}{\mathbf{k}}

\newcommand{\spn}{\mathbf{n}}

\begin{document}
\allowdisplaybreaks[4]
\frenchspacing
\setlength{\unitlength}{1pt}

\title{{\Large\textbf{Inclusive probability to record an electron in elastic \\electromagnetic scattering by a spin one-half hadron wave packet}}}

\date{}

\author{P.O. Kazinski\thanks{E-mail: \texttt{kpo@phys.tsu.ru}}\;, D.I. Rubtsova\thanks{E-mail: \texttt{a5uno@yandex.ru}}\;, and A.A. Sokolov\thanks{E-mail: \texttt{Alexei.sokolov.a@gmail.com}}\\[0.5em]
{\normalsize Physics Faculty, Tomsk State University, Tomsk 634050, Russia}
}
\maketitle

\begin{abstract}

The inclusive probability to record an electron in elastic electromagnetic scattering of an electron by a spin one-half hadron is obtained, the initial quantum states of the electron and the hadron being described by the density matrices of a general form. Contrary to the Rosenbluth formula for the differential cross-section for this process, the first nontrivial contribution to the inclusive probability turns out to be of order $\alpha$ and not $\alpha^2$. This contribution describes the interference between the trivial contribution to the $S$-matrix and the leading contribution to its connected part. The explicit expression for this interference terms is derived. For example, for Gaussian wave packets the interference contribution to the inclusive probability is of relative order $2\times 10^{-2}$ in comparison with the contribution of the free passed particle for the nonrelativistic proton and the $10$ MeV electron with the normal deviation of momenta in the electron and hadron wave packets of order $10$ keV. It is shown that the same interference term arises when the electron is scattered by the classical electromagnetic field produced by the hadron electromagnetic current averaged with respect to the free evolving density matrix of the hadron, even in the case of a single hadron. The interference term describes coherent scattering of the electron by the hadron wave packet and is immune to the quantum recoil experienced by a hadron due to scattering. The effective electron mass operator is found on the mass-shell.

\end{abstract}

\section{Introduction}

The coherent processes are very rare in high energy physics. Usually coherence is lost as the energy of scattered particles increases and even a small quantum recoil leads to decoherence. Nevertheless there is a class of quantum electrodynamical processes where the wave functions of particles interact coherently with other particles even at large energies \cite{KazSol22}. For such processes the form of the wave packet of a particle strongly affects the observables even in the case when only a single particle in the state described by this wave packet participates in each individual experiment \cite{KazSol22,MarcuseII,Talebi16,PanGov18,GovPan18,PanGov19,pra103,PanGov21}. For coherent scattering by a particle wave packet, the scattering amplitudes stemming from different parts of the wave packet add up constructively and so one may regard the wave packet as some charged gas or a fluid. For incoherent scattering by a particle wave packet, the scattering amplitudes from different parts of the wave packet add up destructively. As a result, in a certain approximation, the scattering probabilities rather than the amplitudes from different parts of the wave packet are summed in this case \cite{MarcuseI,PMHK08}.

One of the manifestations of coherent scattering is the presence of new resonances as though a particle is scattered by a beam of particles modulated in an appropriate way rather than by a wave packet of a single particle. In that case the modulation is realized on the scale of the particle wave packet that allows one to achieve a much less scale of variation of the scatterer than for a modulation of particle beams and, consequently, to gain resonances at higher energies \cite{Itatani04}. Furthermore, the dependence of observables on a detailed structure of the wave packet of a particle participating in the process enables one to restore its wave function, i.e., it provides one of the means for tomography of a quantum state \cite{Itatani04,SHDLSL23}. This, in turn, allows one to investigate the influence of the fundamental features of quantum theory, such as entanglement and reduction of a quantum state upon measurement, on the processes in high energy physics.

In the present paper, we investigate coherent electromagnetic scattering of a Dirac fermion (an electron) by a wave packet of the other Dirac fermion (a spin one-half hadron) with the electromagnetic current depending on the two form factors. It turns out that the contribution to the inclusive probability to record an electron arising from the interference of the identical contribution to the $S$-matrix and its connected part describes coherent scattering of this particle by the wave function of the hadron center-of-mass \cite{KazSol22}. This interference contribution is such as if the electron is scattered by a classical electromagnetic field created by the wave function of the hadron even in the case when only a single hadron takes part in each experiment. The quantum recoil due to scattering is absent in this contribution. To put it another way, it is the interference contribution to the inclusive probability to the record the electron that contains the average hadron electromagnetic current discussed in attempts to give a precise physical interpretation to the form factors \cite{Miller19,Lorce20,Jaffe21,FreeMill22,EGLMP22,PEGM22,ASEGM23,ChenLorc23}. The ``standard'' contribution to the inclusive probability proportional to the differential cross-section in the planewave limit includes with necessity the quantum recoil and is determined by the nondiagonal elements of the transition current. As a result, the standard contribution to the inclusive probability is better to interpret as describing incoherent scattering by a particle wave packet \cite{MarcuseI,PMHK08,CorsPeat11,WCCP16,Remez19,KdGdAR21,ADDG22,Wong21,MKMPSY22,LMKMPY23}. Notice that the interference contribution is of lower order in the coupling constant (of order $\alpha$) than the modulus squared of the connected part of the $S$-matrix (of order $\alpha^2$). However, this contribution exists only for non-planewave initial states of scattered particles. In the case when the initial states of particles are plane waves, the interference contribution is different from zero only on the set of points of measure zero in the momentum space \cite{KazSol22}. In order to observe the interference contribution to the inclusive probability, the detector should be placed in front of the emitter of probing particles to record the interference pattern formed on the bright spot produced by the free passed particles. The similar effect for the light-by-light scattering was investigated in \cite{KazSol23}.

The study of the interference contribution to the inclusive probability allows one, in principle, to recover the center-of-mass wave function of the hadron and its form factors. However, in order to retrieve the form factors, it is necessary to prepare such states of the scattered electron that the standard deviation of momenta in these states to be of order of the transferred momentum where the form factor is investigated. As for hadrons such as protons and neutrons, this quantity is of order $1$ GeV and higher. The large dispersions of momenta in the probing electron wave packet can be created, for example, by transverse colliding the electron with the other particle possessing the momentum of order of the normal deviation required. This accessory particle can be the electron as well. In such a case, the whole process runs as follows. At first, the electron wave packet is scattered by the accessory particle to acquire a large dispersion of momenta. Then the escaping electron wave packet is scattered on the hadron wave packet. As far as the shape of the hadron center-of-mass wave function is concerned, the normal deviation of momenta in the probing electron wave packet can be much smaller than $1$ GeV. In this case, the normal deviation just defines the accuracy with which the hadron center-of-mass wave function can be recovered from scattering data.

The paper is organized as follows. In Sec. \ref{GenForm}, the basic definitions are given. In Sec. \ref{Inclus_Probab_Sec}, we evaluate the inclusive probability to record an electron scattered by a hadron wave packet. We derive the general formula and make several approximations in order to simplify it. In Sec. \ref{Mass_Oper_Sec}, we provide a clear physical interpretation to the formulas obtained in the previous section and find the effective mass operator of an electron on the mass-shell. In Conclusion, we summarize the results. In Appendix \ref{App_Traces_Contract}, the traces and the contractions appearing in evaluating the inclusive probability and the mass operator are presented. In Appendix \ref{App_Gauss_wave_p}, the explicit expressions for the contributions to the inclusive probability to record an electron are given in the case of Gaussian initial wave packets of the scattered particles.

We follow the notation adopted in \cite{KazSol22}. The Greek indices $\mu$, $\nu$, and so on are the space-time indices taking the values $\overline{0,3}$ and the Latin $i$, $j$, and so on are the spatial indices. The summation over repeated indices is always understood unless otherwise stated. We also suppose that the quantum states of particles are normalized to unity in some sufficiently large volume $V$. We use the system of units such that $\hbar=c=1$ and $e^2=4\pi\al$, where $\al$ is the fine structure constant. The Minkowski metric is taken with the mostly minus signature.

\section{Basic formulas}\label{GenForm}

The electromagnetic interaction of an electron with a hadron is described by the action functional with the Lagrangian density (see, e.g., \cite{LandLifQED})
\begin{equation}\label{QED_hadron_electron}
	\mathcal{L} = -\frac14 F_{\mu\nu}F^{\mu\nu} +\bar{\psi}_e (i\ga^\mu\partial_\mu-m) \psi_e -\bar{\psi}_e \ga^\mu A_\mu\psi_e
    +\bar{\psi}_n (i\ga^\mu\partial_\mu-M)\psi_n -e\bar{\psi}_n \Ga^\mu A_\mu\psi_n,
\end{equation}
where $m$ is the electron mass, $M$ is the hadron mass, $F_{\mu\nu}:=\partial_{[\mu} A_{\nu]}$ is the electromagnetic stress tensor, $\psi_{e}$ is the Dirac spinor describing the quantum electron-positron field, $\psi_{n}$ is the Dirac spinor describing the quantum hadron field of the spin $1/2$, $\Ga^\mu$ is a nonlocal operator describing the electromagnetic interaction of a hadron in elastic scattering. The action \eqref{QED_hadron_electron} must also be gauge fixed. Henceforth, we imply the Feynman gauge.

We specify the normalization of the Dirac spinors and the mode functions of the electromagnetic field as in \cite{BaKaStrbook,pra103}. In particular,
\begin{equation}
\begin{split}
	\hat\psi_e(x)&=\sum_s\int \frac{V d\spp}{(2\pi)^3} \sqrt{\frac{m}{V p_0}} \Bigl[u^e_s(\spp)e^{-ip_\mu x^\mu}
    \hat{a}_{es}(\spp)+v^e_s(\spp)e^{i p_\mu x^\mu} \hat{b}_{es}^{\dag}(\spp)\Bigr], \\
	\hat\psi_n(x)&=\sum_r\int \frac{V d\spk}{(2\pi)^3} \sqrt{\frac{M}{V k_0}} \Bigl[u^n_r(\spk)e^{-ik_\mu x^\mu}
    \hat{a}_{nr}(\spk) +v^n_r(\spk)e^{i k_\mu x^\mu} \hat{b}_{nr}^{\dag}(\spk)\Bigr],
\end{split}
\end{equation}
where the mass-shell conditions $p_0=\sqrt{m^2+\spp^2}$, $k_0=\sqrt{M^2+\spk^2}$ are fulfilled, $\hat{a}^\dag_{es}(\spp)$, $\hat{a}_{es}(\spp)$ and $\hat{a}^\dag_{nr}(\spk)$, $\hat{a}_{nr}(\spk)$ are the creation and annihilation operators of electrons and hadrons, respectively. The creation and annihilation operators denoted by $\hat{b}^\dag$, $\hat{b}$ are for the corresponding antiparticles. Besides,
\begin{equation}\label{spinors}
\begin{aligned}
	u^e_s(\spp)&=\frac{m+\hat p}{\sqrt{2m(p_0+m)}} \genfrac{[}{]}{0pt}{}{\chi_s}{0},&\qquad v^e_s(\spp)
    &=\frac{m-\hat p}{\sqrt{2m(p_0+m)}} \genfrac{[}{]}{0pt}{}{0}{\chi_s},\\
	u^n_r(\spk)&=\frac{M+\hat k}{\sqrt{2M(k_0+M)}} \genfrac{[}{]}{0pt}{}{\chi_r}{0},&\qquad v^n_r(\spk)
    &=\frac{M-\hat k}{\sqrt{2M(k_0+M)}} \genfrac{[}{]}{0pt}{}{0}{\chi_r},
\end{aligned}
\end{equation}
where the hat over a $4$-vector means the contraction of this $4$-vector with $\ga_\mu$, the standard representation of the $\ga$ matrices is implied, and
\begin{equation}
	(\bs\tau \bs\s) \chi_s=s \chi_s,\qquad s=\pm1.
\end{equation}
The unit vector
\begin{equation}
	\bs\tau=(\sin\theta\cos\vf,\sin\theta\sin\vf,\cos\theta)
\end{equation}
specifies the direction of the spin projection quantization axis. There are the relations
\begin{equation}\label{energy_proj}
\begin{aligned}
	\sum_s u^e_s(\spp)\bar u^e_s(\spp)&=\frac{m+ \hat p}{2m},&\qquad\sum_s v^e_s(\spp)\bar v^e_s(\spp)&=-\frac{m-\hat p}{2m},\\
	\sum_r u^n_r(\spk)\bar u^n_r(\spk)&=\frac{M+\hat k}{2M},&\qquad\sum_r v^n_r(\spk)\bar v^n_r(\spk)&=-\frac{M-\hat k}{2M}.
\end{aligned}
\end{equation}
Henceforth, the indices $e$ and $n$ of the spinors $u_s^e$ and $u_r^n$ will be omitted and the species of a spinor will be fixed by its arguments. The spinors with the spin projection indices $s$ or $s'$ and momenta $\spp$ or $\spp'$ describe the electron states, whereas the spinors with indices $r$ or $r'$ and momenta $\spk$ or $\spk'$ are reserved for hadrons. Furthermore, we use the condensed indices and notation. For example,
\begin{equation}
    \al:=(s,\spp),\qquad \sum_\al:=\sum_s \int \frac{Vd\spp}{(2\pi)^3},
\end{equation}
and $\al':=(s',\spp')$, $\be:=(r,\spk)$, $\be':=(r',\spk')$.

The effective hadron electromagnetic vertex on the hadron mass-shell has the form \cite{LandLifQED,Miller19}
\begin{equation}\label{hadron_curr}
    \bar{u}^{r'}(\spk')\Ga^\mu(k',k) u^r(\spk)=\bar{u}^{r'}(\spk')\Big\{[F_e(q^2)-F_m(q^2)] \frac{Mk_c^\mu}{k_c^2}+F_m(q^2) \ga^\mu\Big\} u^r(\spk),
\end{equation}
where $q:=k'-k$ and $k_c:=(k+k')/2$. This matrix element is orthogonal to $q^\mu$. It is also clear that $q_\mu k_c^\mu=0$. The form factor $F_e(q^2)$ characterizes the charge distribution in a hadron, the form factor $F_m(q^2)$ specifies the magnetic moment density, and $F_m(q^2)-F_e(q^2)$ describes the anomalous magnetic moment distribution. The model \eqref{QED_hadron_electron} can also be employed for description of electron muon scattering. In that case, one needs to set $F_e(q^2)=F_m(q^2)=1$ and to take $M$ to be equal to the muon mass.

\section{Inclusive probability}\label{Inclus_Probab_Sec}

Consider elastic electromagnetic scattering of an electron by a hadron. The transition amplitude for this process is determined by the matrix element of the evolution operator
\begin{equation}
	A(\al',\be';\al,\be):=\braket{\al',\be'|\hat U_{t_2,t_1}|\be,\al}=\braket{\al',\be'|\hat{U}^0_{t_2,0}\hat{S}_{t_2,t_1}
    \hat{U}^0_{0,t_1}|\be,\al},
\end{equation}
where $\hat{U}^0_{t_2,t_1}$ is the free evolution operator and
\begin{equation}\label{smatrix}
	\hat{S}_{t_2,t_1}=\Texp\Big\{-i\int_{t_1}^{t_2} dx \Big[:e\hat{\bar\psi}^e\ga^\mu \hat A_\mu\hat\psi^e:
    +:e\hat{\bar\psi}^n\Ga^\mu \hat A_\mu \hat\psi^n :\Big]\Big\}.
\end{equation}
Using the relations
\begin{equation}
	\hat{U}^0_{0,t}\hat{a}_\al \hat{U}^0_{t,0}=e^{-iE_\al t} \hat{a}_{\al},\qquad \hat{U}^0_{t_2,t_1}\ket{0}=e^{-iE_0(t_2-t_1)}\ket{0},
\end{equation}
where $E_\al$ is the energy of the one-particle state $\al$ and $E_0$ is the vacuum energy, the matrix element is reduced to
\begin{equation}
	A(\al',\be';\al,\be)=e^{-iE_0(t_2-t_1)}e^{i(E_\al+E_\be)t_1}e^{-i(E_{\al'}+E_{\be'})t_2}\braket{\al',\be'|\hat S_{t_2,t_1}|\be,\al}.
\end{equation}
We take the initial states of the electron and the hadron at the instant of time $t_1$ to be the wave packets of a general form
\begin{equation}\label{vp}
\begin{gathered}
	\ket{\phi}=\sum_\al \sqrt{\frac{(2\pi)^3}{V}} \tilde{\phi}(\al) \ket{\al},\qquad \ket{\psi}=\sum_\be \sqrt{\frac{(2\pi)^3}{V}}
    \tilde{\psi}(\be) \ket{\be}, \\
	\sum_s \int d\spp |\tilde{\phi}_s(\spp)|^2=1,\qquad\sum_r \int d\spk |\tilde{\psi}_r(\spk)|^2=1.
\end{gathered}
\end{equation}
In that case, the transition amplitude is written as
\begin{equation}
	A(\al',\be';\phi,\psi)=\sum_{\al,\be}\frac{(2\pi)^3}{V}	A(\al',\be';\al,\be) \tilde{\phi}(\al)\tilde{\psi}(\be).
\end{equation}
It is useful to specify the forms of the wave packets at the instant of time $t=0$ instead of $t=t_1$. Then
\begin{equation}
	\phi(\al):=e^{iE_\al t_1}\tilde{\phi}(\al),\qquad \psi(\be):=e^{iE_\be t_1}\tilde{\psi}(\be).
\end{equation}
It is clear that the wave functions $\phi(\al)$ and $\psi(\be)$ satisfy the same normalization conditions \eqref{vp}.

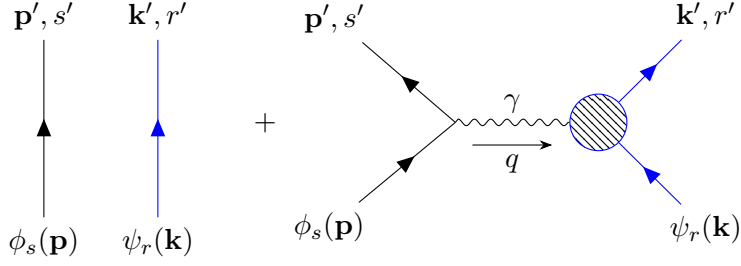
\begin{figure}[t]
	 \center{
	\begin{tikzpicture}
		\begin{feynman}
			\vertex (i1) {$\phi_s(\spp)$};
			\vertex [above =3cm of i1] (i2) {$\spp',s'$};
			\vertex [right =of i1] (i3) {$\psi_r(\spk)$};
			\vertex [above =3cm of i3] (i4) {$\spk',r'$};
			\vertex [below right =2cm of i4] (c) {$+$};
			\vertex [right =2.5cm of c](f1);
			\node [blue,blob, right =of f1](f2);
			\vertex [below left=of f1] (k1) {$\phi_s(\spp)$};
			\vertex [below right=2cm of f2] (k2) {$\psi_r(\spk)$};
			\vertex [above left=of f1] (k3) {$\spp',s'$};
			\vertex [above right=2cm of f2] (k4) {$\spk',r'$};
			\diagram{
				(i1) -- [fermion] (i2);
				(i3) -- [blue,fermion] (i4);	
				(k1) -- [fermion] (f1);
				(k2) -- [blue,fermion] (f2);
				(f2) -- [boson, edge label'=$\ga$, reversed momentum=$q$] (f1);
				(f1) -- [fermion] (k3);
				(f2) -- [blue,fermion] (k4);
			};
		\end{feynman}
	\end{tikzpicture}
	}	
\caption{{\footnotesize The diagrams describing electron hadron scattering in the leading order of the perturbation theory. The time axis is directed upwards. The blue lines correspond to hadrons, whereas the black ones are for electrons.}}\label{ris:FeynDiag}
\end{figure}

Developing the $T$-exponent in \eqref{smatrix} as a series up to the terms quadratic in $e$, discarding the irrelevant phase factors, and going to the limits $t_1\rightarrow-\infty$, $t_2\rightarrow\infty$, we arrive at the matrix element of the evolution operator
\begin{equation}\label{apk2}
\begin{split}
	A(\al',\be';\phi,\psi)=\,&\frac{(2\pi)^3}{V}\Big\{\phi_{s'}(\spp')\psi_{r'}(\spk')+\\
    &+\frac{ie^2mM}{(2\pi)^2}\sum_{r,s}\int d\spp
    d\spk\phi_s(\spp)\psi_r(\spk)\frac{\de(p'+k'-p-k)}{q^2+i\e}\frac{\bar u^{s'}(\spp')\ga^\mu u^s(\spp)\, \bar u^{r'}(\spk')\Ga_\mu u^r(\spk)}{\sqrt{p'_0k'_0p_0k_0}}\Big\}.
\end{split}
\end{equation}
The sum of the two terms given above corresponds to the Feynman diagrams depicted in Fig. \ref{ris:FeynDiag}.

Let us find the inclusive probability to record the electron after scattering with momenta $(\spp',\spp'+d\spp')$ with a certain polarization vector $\bs\zeta$. Such an inclusive probability reads
\begin{equation}\label{inclusive_prob_defn}
	dP(\bs\zeta,\spp')=\sum_{s'_1,s'_2} D_{s'_2s'_1}(\bs\zeta)w_{s'_1s'_2}(\spp') d\spp',
\end{equation}	
where
\begin{equation}\label{w_defn}
    w_{s'_1s'_2}(\spp'):=\frac{V}{(2\pi)^3} \sum_{r'}\int \frac{V d\spk'}{(2\pi)^3} A(\spp',s'_1;\spk',r';\phi,\psi)
    A^*(\spp',s'_2;\spk',r';\phi,\psi)
\end{equation}
is the electron density matrix after scattering and the projector
\begin{equation}
    D_{s'_2s'_1 }(\bs\zeta):=\frac12[1+(\bs\s\bs\zeta)]_{s'_2s'_1}
\end{equation}
singles out the electron state with polarization $\bs\zeta$. If the density matrix $w_{s'_1s'_2}$ is expanded in terms of the Pauli matrices,
\begin{equation}
    w_{s'_1s'_2}=\frac{\rho_e'}{2}\big[\de_{s'_1s'_2}+(\xi'_e)_k(\s_k)_{s'_1s'_2}\big],
\end{equation}
then
\begin{equation}\label{inclusive_prob_gen}
    dP(\bs\zeta,\spp')=\frac{\rho_e'}{2}[1+(\bs{\zeta}\bs{\xi}'_e)] d\spp'.
\end{equation}
Further we will find the explicit expressions for $\rho'_e$ and $\bs{\xi}'_e$ (see formulas \eqref{w_expanded}, \eqref{w_expanded_1}).

Introduce the density matrices of initial states of the electron and the hadron
\begin{equation}\label{dens_matrices}
    \rho^e_{ss'}(\spp,\spp')=\frac{\rho_e(\spp,\spp')}{2}[1+(\bs\s\bs{\xi}_e(\spp,\spp'))]_{ss'},\qquad \rho^n_{rr'}(\spk,\spk')=\frac{\rho_n(\spk,\spk')}{2}[1+(\bs\s\bs{\xi}_n(\spk,\spk'))]_{rr'}.
\end{equation}
It is clear that
\begin{equation}
    \rho^*_e(\spp,\spp')=\rho_e(\spp',\spp),\qquad \bs{\xi}^*_e(\spp,\spp')=\bs{\xi}_e(\spp',\spp).
\end{equation}
The same relations take place for the hadron density matrix. For pure states,
\begin{equation}\label{pure_states}
    \rho^e_{ss'}(\spp,\spp')=\phi_{s}(\spp)\phi^*_{s'}(\spp'),\qquad \rho^n_{rr'}(\spk,\spk')=\psi_{r}(\spk)\psi^*_{r'}(\spk').
\end{equation}
Then $w_{s'_1s'_2}$ is the sum of four terms
\begin{equation}\label{w_contribs}
\begin{split}
    w^{(1)}_{s'_1s'_2}=\,&\rho^e_{s'_1s'_2}(\spp',\spp'),\\
    w^{(2)}_{s'_1s'_2}=\,&\frac{ie^2mM}{(2\pi)^2}\sum_{r',r,s}\int d\spp d\spk d\spk' \rho^e_{ss'_2}(\spp,\spp')\rho^n_{rr'}(\spk,\spk') \frac{\de(p'+k'-p-k)}{q^2+i0}\times\\
	&\times\frac{\bar u^{s'_1}(\spp')\ga^\mu u^s(\spp)\, \bar u^{r'}(\spk') \Ga_\mu
    u^r(\spk)}{\sqrt{p'_0k'_0p_0k_0}},\\
    w^{(3)}_{s'_1s'_2}=\,&\frac{-ie^2mM}{(2\pi)^2}\sum_{r',r,s}\int d\spp d\spk d\spk' [\rho^e_{ss'_1}(\spp,\spp')]^* [\rho^n_{rr'}(\spk,\spk')]^* \frac{\de(p'+k'-p-k)}{q^2-i0}\times\\
	&\times\frac{\big[\bar u^{s'_2}(\spp')\ga^\mu u^s(\spp)\big]^*\big[ \bar u^{r'}(\spk')\Ga_\mu
    u^r(\spk)\big]^*}{\sqrt{p'_0k'_0p_0k_0}},\\
    w^{(4)}_{s'_1s'_2}=\,&\frac{e^4 m^2 M^2}{(2\pi)^4}\sum_{r',r,s,\tilde r,\tilde s}\int d\spp d\spk d\tilde{\spp} d\tilde{\spk} d\spk' \rho^e_{s\tilde{s}}(\spp,\tilde{\spp}) \rho^n_{r\tilde{r}}(\spk,\tilde{\spk}) \frac{\de(p'+k'-p-k)}{q^2+i0}
	\frac{\de(p'+k'-\tilde p-\tilde k)}{\tilde q^2-i0}\times\\
	&\times\frac{\bar u^{s'_1}(\spp')\ga^\mu u^s(\spp)\,\bar u^{r'}(\spk')\Ga_\mu u^r(\spk)\,\bar u^{\tilde s}
    (\tilde \spp)\ga^\nu u^{s'_2}(\spp')\, \bar u^{\tilde r}(\tilde \spk)\Ga_\nu u^{r'}(\spk')}{p'_0k'_0\sqrt{p_0k_0\tilde p_0\tilde k_0}},
\end{split}
\end{equation}
where $\tilde{q}:=k'-\tilde{k}$. It is evident that
\begin{equation}
    [w^{(2)}_{s'_1s'_2}]^*=w^{(3)}_{s'_2s'_1}.
\end{equation}
The fourth term, $w^{(4)}_{s'_1s'_2}$, gives rise to the standard contribution to the differential cross-section of elastic electromagnetic scattering of an electron by a hadron, the nontrivial form of wave packets of the electron and the hadron being taken into account. This contribution is of order $\al^2$. Employing relations \eqref{energy_proj}, we bring this term into the form
\begin{equation}\label{w4}
\begin{split}
    w^{(4)}_{s'_1s'_2}= \,&\frac{e^4 m^2 M}{(2\pi)^4}\sum_{r,s,\tilde r,\tilde s}\int d\spp d\spk d\tilde{\spp} d\tilde{\spk} d\spk' \rho^e_{s\tilde{s}}(\spp,\tilde{\spp}) \rho^n_{r\tilde{r}}(\spk,\tilde{\spk}) \frac{\de(p'+k'-p-k)}{q^2+i0}
	\frac{\de(p'+k'-\tilde p-\tilde k)}{\tilde q^2-i0}\times\\
	&\times\frac{\bar u^{s'_1}(\spp')\ga^\mu u^s(\spp)\, \bar u^{\tilde s}
    (\tilde \spp)\ga^\nu u^{s'_2}(\spp')\,\bar u^{\tilde r}(\tilde \spk)\Ga_\nu (M+\hat{k}')\Ga_\mu u^r(\spk)  }{2p'_0k'_0\sqrt{p_0k_0\tilde p_0\tilde k_0}}.
\end{split}
\end{equation}
In the case when the initial states of the electron and the hadron are plane waves, this contribution turns into the Rosenbluth formula for the differential cross-section \cite{Rosen50,LandLifQED}. Notice that expression \eqref{w4} does not depend on the diagonal matrix element of the hadron electromagnetic current \eqref{hadron_curr} (see Sec. \ref{Mass_Oper_Sec}) but it depends on the sum of moduli squared of the transition current with the different final momenta $\spk'$ and the spin projections $r'$. Therefore, in interpreting this contribution to the inclusive probability, it is faulty to think of the electron scattered by the hadron wave packet as if the electron is scattered by some gas of charged particles with magnetic moments whose density is determined by the wave function of the hadron center-of-mass (see, e.g., \cite{MarcuseI,PMHK08,CorsPeat11,WCCP16,Remez19,KdGdAR21,ADDG22,Wong21,MKMPSY22,LMKMPY23}). It should also be noted that the term \eqref{w4} is not the only contribution to the inclusive probability of order $\al^2$. There are other contributions of this order stemming from the loop corrections. The investigation of these terms will be given elsewhere.

In what follows, we will be interested in the second and third contributions to the density matrix $w_{s'_1s'_2}$. These terms stem from the interference of the diagrams presented in Fig. \ref{ris:FeynDiag}. These contributions are linear in $\al$ and are absent in the planewave approximation. In the planewave approximation, they are different from zero only on the zero measure set in the momentum space \cite{KazSol22}. In the domain of the electron momenta $\spp'$, where $\rho_e(\spp',\spp')$ is not small, these terms are much larger than the contribution $w^{(4)}_{s'_1s'_2}$ provided $Z\al\ll1$, where $Z$ is the hadron charge, i.e., in the case where the standard perturbation theory is valid. By the order of magnitude, the interference terms dominate over the $\alpha^2$ term provided $|w^{(2)}_{s'_1s'_2}+w^{(3)}_{s'_1s'_2}|/|w^{(1)}_{s'_1s'_2}|\ll1$. The explicit expressions for this estimate will be given at the end of this section. Furthermore, as is seen from \eqref{w_contribs}, the contributions proportional to $\alpha$ are determined by the expectation value of the hadron electromagnetic current in the free evolved state of the hadron center-of-mass. In that sense, these contributions are immune to the quantum recoil caused by the interaction of the electron with the hadron (see for details \cite{KazSol22,pra103}).

It is convenient to reduce the evaluation of the current matrix elements appearing in $w^{(2,3)}_{s'_1s'_2}$ to the evaluation of traces of products of the $\ga$ matrices. As follows from the explicit form of the spinors \eqref{spinors}, the matrix element
\begin{equation}
	\bar u^{s'}(\spp')\ga^\mu u^s(\spp)=\frac{\Sp\Big[\ga^\mu(m+\hat p)
    \left[
      \begin{array}{cc}
        \chi_s\chi_{s'}^\dag & 0 \\
        0 & 0 \\
      \end{array}
    \right]
    (m+\hat p')\Big]}{2m\sqrt{(p_0+m)(p'_0+m)}}.
\end{equation}
The product of the spinors characterizing the spin of the electron state can be expanded in terms of the Pauli matrices
\begin{equation}
	\chi^\varsigma_s\,(\chi^{\varsigma'}_{s'})^*=\frac12\big[\delta_{s's}\de^{\varsigma\varsigma'}
    +\tau^i_a(\sigma^i)^{\varsigma\varsigma'} (\sigma_a)_{s's}\big],
\end{equation}
where
\begin{equation}
\begin{gathered}
    \bs{\tau}_1=\re\mathbf{f},\qquad \bs{\tau}_2=\im\mathbf{f},\qquad\bs{\tau}_3=\bs\tau,\\
    \mathbf{f}=e^{i\vf}(\cos\theta \cos\vf-i\sin\vf,\cos\theta \sin\vf+i\cos\vf,-\sin\theta).
\end{gathered}
\end{equation}
Notice that, formally, the vector $\mathbf{f}$ is proportional to the polarization vector of a right-handed photon propagating along the direction $\bs\tau$. The set of vectors $\tau^i_a$ can be regarded as a right-handed tetrad mapping the vector on the Poincar\'{e} sphere to the spin vector in the $x$-space. Bearing in mind that
\begin{equation}
    \left[
      \begin{array}{cc}
        \s^i & 0 \\
        0 & 0 \\
      \end{array}
    \right]=-\ga^i \ga^5\frac{1+\ga^0}{2},
\end{equation}
we come to
\begin{equation}
    \bar u^{s'}(\spp')\ga^\mu u^s(\spp)=\frac{\Sp\Big[\ga^\mu(m+\hat p)
    [\de_{s's} +(\s_a)_{s's}\tau_{ai}\ga^i\ga^5]\frac{1+\ga^0}{2}
    (m+\hat p')\Big]}{4m\sqrt{(p_0+m)(p'_0+m)}}.
\end{equation}
The traces of the respective products of the $\ga$ matrices are calculated in Appendix \ref{App_Traces_Contract}. Employing the notation introduced there, we deduce
\begin{equation}\label{electron_curr_matr_elem}
    \bar u^{s'}(\spp')\ga^\mu u^s(\spp)=\frac12[\de_{s's}G^\mu(\spp,\spp')-(\s_a)_{s's}\tau_{ai}Z^{i\mu}(\spp,\spp')].
\end{equation}
Analogously, for the matrix element of the hadron electromagnetic current, we have
\begin{equation}\label{hadron_curr_matr_elem}
    \bar u^{r'}(\spk')\Ga^\mu u^r(\spk)=\frac12[\de_{r'r}\tilde{G}^\mu(\spk,\spk')-(\s_a)_{r'r}\tau_{ai}\tilde{Z}^{i\mu}(\spk,\spk')].
\end{equation}
As in the case of electrons, the quantization axis of spin projection is directed along the vector $\bs\tau$.

Substituting \eqref{dens_matrices}, \eqref{electron_curr_matr_elem}, and \eqref{hadron_curr_matr_elem} into \eqref{w_contribs} and taking into account the properties \eqref{rels}, \eqref{tilde_G_prop}, we arrive at
\begin{equation}\label{inclusive_prob}
\begin{split}
    w^{(2)}_{s'_1s'_2}+w^{(3)}_{s'_1s'_2}=\,&- \frac{e^2mM}{4(2\pi)^2}\int d\spp d\spk d\spk'\frac{\de(p'+k'-p-k)}{q^2\sqrt{p_0p'_0k_0k'_0}} \Big\{\im\Big[\rho_e\rho_n \big(\tilde{G}_n^\mu G^e_\mu -\xi^a_e \tau_{ai}Z^{i\mu}_e\tilde{G}^n_\mu-\\
    &-\xi^a_n\tau_{ai}\tilde{Z}^{i\mu}_nG^e_\mu +\xi^a_n\tau_{ai}\tilde{Z}^{i\mu}_n Z^e_{j\mu} \tau^j_b \xi^b_e\big) \Big]\de_{s'_1s'_2} +\im\Big[\rho_e\rho_n \big(\xi^k_e\tilde{G}_n^\mu G^e_\mu
    -\xi^k_e\xi^a_n \tau_{ai}\tilde{Z}^{i\mu}_nG^e_\mu-\\
    &-\big(\de_{ak} -i\xi^i_e\e_{iak}\big)\tau_{aj}Z^{j\mu}_e\tilde{G}^n_\mu+\tau_a^l Z^e_{l\mu}\tilde{Z}_n^{j\mu}\tau_{bj}\xi^b_n \big(\de_{ak} +i\xi^i_e\e_{aik} \big) \big) \Big](\s_k)_{s'_1s'_2} \Big\}.
\end{split}
\end{equation}
In virtue of the fact that $q^2=(k'-k)^2\leq0$ and $q^2=0$ if and only if $k'=k$, the integrands of integrals in $w^{(2,3)}_{s'_1s'_2}$ do not contain nonintegrable singularities. Therefore, the additions $\pm i0$ in the photon propagators can be thrown away. The explicit expressions for the contractions appearing in \eqref{inclusive_prob} are given in Appendix \ref{App_Traces_Contract}. Formula \eqref{inclusive_prob} takes the form
\begin{equation}
    w^{(2)}_{s'_1s'_2}+w^{(3)}_{s'_1s'_2}=\frac12[a\de_{s'_1s'_2}+b_k(\s_k)_{s'_1s'_2}].
\end{equation}
Hence
\begin{equation}\label{w_expanded}
    w^{(1)}_{s'_1s'_2}+w^{(2)}_{s'_1s'_2}+w^{(3)}_{s'_1s'_2}\approx\frac{\rho'_e}{2}[\de_{s'_1s'_2}+({\bs\s}\bs{\xi}'_e)_{s'_1s'_2}],
\end{equation}
where
\begin{equation}\label{w_expanded_1}
    \rho'_e=\rho_e+a,\qquad \bs{\xi}'_e=\bs{\xi}_e+\mathbf{b}/\rho_e.
\end{equation}
The last expression describes a change of the electron polarization vector in scattering by a hadron wave packet.

Consider several particular cases of the general formula \eqref{inclusive_prob}. It follows from this formula that $|\spq|$ is of order of the minimum of the standard deviations of momenta in the wave packets of the electron and the hadron. Suppose that the dispersions of momenta in the electron and hadron wave packets are such that the estimates \eqref{small_recoil} are fulfilled. Consider the two cases: a) the hadron is electrically charged, viz., $|F_e|\sim |F_m| \gtrsim1$; b) the hadron is electrically neutral (the neutron), viz., $|F_e|\sim |q^2|/M^2$, $|F_m|\gtrsim1$. Notice that electron muon scattering comes into the case (a) by setting $F_e=F_m=1$.

In the case (a), we have
\begin{equation}\label{w23_electr}
    w^{(2)}_{s'_1s'_2}+w^{(3)}_{s'_1s'_2}=-e^2\int \frac{d\spk' d\spp}{(2\pi)^2} \frac{\de(p_0'+k_0'-p_0-k_0)}{(p'-p)^2 p'_0k'_0} (p'k')F_e \big((p'-p)^2\big) \big[\im(\rho_e\rho_n)\de_{s'_1s'_2} +\im(\rho_e\rho_n\xi^k_e)(\s_k)_{s'_1s'_2} \big],
\end{equation}
where $\spk=\spk'+\spp'-\spp$. The argument of the delta function expressing the energy conservation law can be simplified
\begin{equation}
    p_0'+k_0'-p_0-k_0\approx -(\De\bs\be \spq),
\end{equation}
where $\De\bs\be:=\bs{\be}'_e-\bs{\be}'_n$ and
\begin{equation}
    \bs{\be}'_e:=\spp'/p_0',\qquad \bs{\be}'_n:=\spk'/k_0'.
\end{equation}
Let us split the momentum components into longitudinal and perpendicular to the vector $\De\bs\be$, for example,
\begin{equation}
    \spq=\spq_\perp +q_\parallel\De\bs\be/|\De\bs\be|,\qquad q_\parallel:=\frac{(\De\bs\be \spq)}{|\De\bs\be|}.
\end{equation}
Then
\begin{equation}
    \de(p_0'+k_0'-p_0-k_0)\approx \de(q_\parallel)/|\De\bs\be|.
\end{equation}
Assume additionally that the wave packet of the hadron center-of-mass is such that $\De\bs\be$ can be regarded as constant. This is the case, for example, for a nonrelativistic hadron and a relativistic electron or for a wave packet of the hadron center-of-mass with small deviation of momenta in comparison with the average momentum. As a result,
\begin{equation}\label{w23_parax}
    w^{(2)}_{s'_1s'_2}+w^{(3)}_{s'_1s'_2}=-\frac{e^2}{|\De\bs\be|} \int \frac{d\spk' d\spq_\perp}{(2\pi)^2} (1-(\bs{\be}'_e\bs{\be}'_n)) \frac{F_e\big((\bs{\be}'_e\spq_\perp)^2-\spq_\perp^2\big)}{(\bs{\be}'_e\spq_\perp)^2-\spq_\perp^2} \big[\im(\rho_e\rho_n)\de_{s'_1s'_2} +\im(\rho_e\rho_n\xi^k_e)(\s_k)_{s'_1s'_2} \big],
\end{equation}
where
\begin{equation}\label{notat_near_diag}
    \rho_n=\rho_n(\spk'-\spq_\perp,\spk'), \qquad\rho_e=\rho_e(\spp'+\spq_\perp,\spp'),\qquad \xi_e=\xi_e(\spp'+\spq_\perp,\spp').
\end{equation}
If the hadron is nonrelativistic and $|\bs{\be}'_n|\ll|\bs{\be}'_e|$ , then we have
\begin{equation}\label{w23_nonrel}
    w^{(2)}_{s'_1s'_2}+w^{(3)}_{s'_1s'_2}=-\frac{e^2}{|\bs{\be}'_e|} \int \frac{d\spk' d\spq_\perp}{(2\pi)^2} \frac{F_e\big(-\spq_\perp^2\big)}{-\spq_\perp^2} \big[\im(\rho_e\rho_n)\de_{s'_1s'_2} +\im(\rho_e\rho_n\xi^k_e)(\s_k)_{s'_1s'_2} \big].
\end{equation}
Notice that this integral is well defined near $\spq_\perp=0$ since the expression in the square brackets in the integrand vanishes for $\spq_\perp=0$. The explicit expression for the interference contribution \eqref{w23_nonrel} in the case of unpolarized initial wave packets of Gaussian profile is presented in Appendix \ref{App_Gauss_wave_p}. Expression \eqref{w23_nonrel} can be cast into the form
\begin{equation}\label{w23_formf}
    w^{(2)}_{s'_1s'_2}+w^{(3)}_{s'_1s'_2}=-\frac{e^2}{|\bs{\be}'_e|} \int \frac{d\spq_\perp}{(2\pi)^2} \frac{F_e\big(-\spq_\perp^2\big)}{-\spq_\perp^2}  \big[\im\big(\tilde{\rho}_n(\spq_\perp) \rho_e\big) \de_{s'_1s'_2} +\im\big(\tilde{\rho}_n(\spq_\perp) \rho_e\xi^k_e\big)(\s_k)_{s'_1s'_2}\big],
\end{equation}
where we have introduced the transverse form factor of the modulus squared of the hadron wave function in the $x$-representation
\begin{equation}
    \tilde{\rho}_n(\spq_\perp):=\int d\spx e^{i\spq_\perp\spx}\rho_n(\spx,\spx).
\end{equation}
If the normal deviation of momenta in the electron wave packet is much smaller than the typical scale of variation of the hadron wave packet in the momentum representation, then
\begin{equation}\label{rho_n_diag}
    \rho_n(\spk'-\spq_\perp,\spk')\approx \rho_n(\spk',\spk'),
\end{equation}
and so
\begin{equation}\label{w23_nonrel_wo_rn}
    w^{(2)}_{s'_1s'_2}+w^{(3)}_{s'_1s'_2}=-\frac{e^2}{|\bs{\be}'_e|} \int \frac{d\spq_\perp}{(2\pi)^2} \frac{F_e\big(-\spq_\perp^2\big)}{-\spq_\perp^2} \big[\im\rho_e\de_{s'_1s'_2} +\im(\rho_e\xi^k_e)(\s_k)_{s'_1s'_2} \big].
\end{equation}
We see that in this case the dependence of the inclusive probability on the form of the hardon wave packet disappears. Formulas \eqref{w23_electr}, \eqref{w23_parax}, \eqref{w23_nonrel}, and \eqref{w23_nonrel_wo_rn} give, in particular, a simple expression for the vector $b_k$ characterizing a change of the electron polarization in scattering by a hadron wave packet.

Now consider the particular case (b), i.e., elastic electromagnetic scattering of an electron by a neutron. In the leading order in $q^\mu$, we obtain
\begin{equation}\label{w23_neutr}
\begin{split}
    w^{(2)}_{s'_1s'_2}+w^{(3)}_{s'_1s'_2}=\,&-\frac{e^2}{2} \int \frac{d\spk' d\spp}{(2\pi)^2} \frac{\de(p_0'+k_0'-p_0-k_0)}{(p'-p)^2 p'_0k'_0} \frac{F_m \big((p'-p)^2\big)}{k'_0+M} (\e^{i\mu\nu\rho}p'_\mu q_\nu (k'_\rho+M\de^0_\rho) -\frac{(p'k')}{M}\e^{0i\nu\rho}q_\nu k'_\rho )\times\\
    &\times\tau_{ai}\big[\re(\rho_e\rho_n\xi^a_n)\de_{s'_1s'_2} +\re(\rho_e\rho_n\xi^a_n\xi^k_e)(\s_k)_{s'_1s'_2} \big],
\end{split}
\end{equation}
where $\spk=\spk'+\spp'-\spp$. As in the case of a charged hadron, for a nonrelativistic neutron we come to
\begin{equation}\label{w23_neutr_nonrel}
    w^{(2)}_{s'_1s'_2}+w^{(3)}_{s'_1s'_2}=\frac{e^2}{2M} \tau_{ai}\e^{ijl}\frac{(\be'_{e})_j}{|\bs{\be}'_e|} \int \frac{d\spk' d\spq_\perp}{(2\pi)^2} \frac{F_m\big(-\spq_\perp^2\big)}{-\spq_\perp^2}  q_{\perp l} \big[\re(\rho_e\rho_n\xi^a_n)\de_{s'_1s'_2} +\re(\rho_e\rho_n\xi^a_n\xi^k_e)(\s_k)_{s'_1s'_2} \big],
\end{equation}
where the notation \eqref{notat_near_diag} has been used and
\begin{equation}
    \xi^a_n=\xi^a_n(\spk'-\spq_\perp,\spk').
\end{equation}
Introducing the transverse form factor of the neutron spin density,
\begin{equation}
    \tilde{\xi}^a_n(\spq_\perp):=\int d\spx e^{i\spq_\perp\spx}\rho_n(\spx,\spx)\xi^a_n(\spx,\spx),
\end{equation}
expression \eqref{w23_neutr_nonrel} can be rewritten as
\begin{equation}\label{w23_neutr_formf}
    w^{(2)}_{s'_1s'_2}+w^{(3)}_{s'_1s'_2}=\frac{e^2}{2M} \tau_{ai}\e^{ijl}\frac{(\be'_{e})_j}{|\bs{\be}'_e|} \int \frac{d\spq_\perp}{(2\pi)^2} \frac{F_m\big(-\spq_\perp^2\big)}{-\spq_\perp^2}  q_{\perp l} \big[\re\big(\tilde{\xi}^a_n(\spq_\perp)\rho_e\big)\de_{s'_1s'_2} +\re\big(\tilde{\xi}^a_n(\spq_\perp)\rho_e\xi^k_e\big) (\s_k)_{s'_1s'_2} \big].
\end{equation}
If the normal deviation of momenta in the electron wave packet is much smaller than the typical scale of variation of the hadron wave packet in the momentum space such that the approximate equality \eqref{rho_n_diag} holds and
\begin{equation}
    \xi_n(\spk'-\spq_\perp,\spk')\approx \xi_n(\spk',\spk'),
\end{equation}
then we have
\begin{equation}\label{w23_neutr_parax}
    w^{(2)}_{s'_1s'_2}+w^{(3)}_{s'_1s'_2}=\frac{e^2}{2M} \lan\xi^a_n\ran \tau_{ai}\e^{ijl}\frac{(\be'_{e})_j}{|\bs{\be}'_e|} \int \frac{d\spq_\perp}{(2\pi)^2} \frac{F_m\big(-\spq_\perp^2\big)}{-\spq_\perp^2}  q_{\perp l} \big[\re \rho_e\de_{s'_1s'_2} +\re(\rho_e\xi^k_e)(\s_k)_{s'_1s'_2} \big],
\end{equation}
where
\begin{equation}
    \lan\xi^a_n\ran:=\int d\spk\rho_n(\spk,\spk)\xi^a_n(\spk,\spk).
\end{equation}
The dependence on the shape of the hadron wave packet disappears in this case. Formulas \eqref{w23_formf}, \eqref{w23_neutr_formf} can be employed for tomography of the modulus squared of the hadron wave function and of the spin structure of its state.

A concise form of the condition when the interference terms are much larger than the standard contribution can be inferred from formulas \eqref{w23_nonrel_wo_rn}, \eqref{w23_neutr_parax}. Formula \eqref{w23_nonrel_wo_rn} implies that the interference terms dominate for scattering of an electron by a charged nonrelativistic hadron provided
\begin{equation}\label{estim_charged}
    \al Z|\rho_e|/|\bs{\be}'_e\rho_e(\spp',\spp')|\ll1,
\end{equation}
where a typical value of $|\rho_e|$ should be taken, for example, the maximum value (see, e.g., Appendix \ref{App_Gauss_wave_p}). As for scattering of an electron by a nonrelativistic neutron, formula \eqref{w23_neutr_parax} gives rise to the estimate
\begin{equation}\label{estim_neutr}
    \al |\rho_e| \sigma_\perp/(M \rho_e(\spp',\spp'))\ll1,
\end{equation}
where $\sigma_\perp$ is the normal deviation of momenta in the electron wave packet. It is clear from the estimates \eqref{estim_charged}, \eqref{estim_neutr} that the interference terms become negligible for those electron momenta $\spp'$ where $\rho_e(\spp',\spp')$ tends to zero.

\section{Mass operator of an electron}\label{Mass_Oper_Sec}

In order to obtain a clear physical interpretation of the formulas obtained in the previous section, consider the action functional of Dirac spinors with the effective self-interaction
\begin{equation}
	S[\bar{\psi},\psi]=\int dx \bar\psi_e(x)(i \ga^\mu \partial_\mu-m)\psi_e(x)-\int dxdy \bar\psi_e(x)M(x,y)\psi_e(y),
\end{equation}
where $M(x,y)$ is the kernel of the mass operator. This operator is nonlocal, in general, and is taken into account perturbatively.

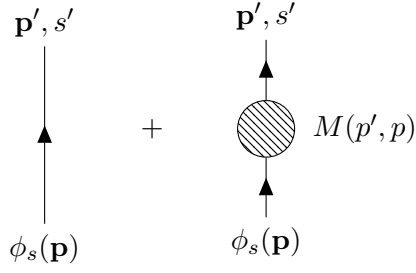
\begin{figure}[t]
\center{
\begin{tikzpicture}
	\begin{feynman}
		\vertex (i1) {$\phi_s(\spp)$};
		\vertex [above =3cm of i1] (i2) {$\spp',s'$};
		\vertex [below right =2cm of i2] (c) {$+$};
		\node [blob, right =1.5cm of c] (b1);
		\vertex [right =1.3cm of b1] (m1) {$M(p',p)$};			
		\vertex [below =of b1] (f1) {$\phi_s(\spp)$};
		\vertex [above =of b1] (f2) {$\spp',s'$};
		\diagram{
		(i1) -- [fermion] (i2);
		(f1) -- [fermion] (b1);
		(b1) -- [fermion] (f2);
		};
	\end{feynman}
\end{tikzpicture}
}
\caption{{\footnotesize The diagrams describing electron scattering by the effective potential determined by the mass operator in the leading order of perturbation theory.}}\label{ris:FeynDiag2}
\end{figure}

The transition amplitude in the leading order of perturbation theory becomes
\begin{equation}\label{amplitude_mass_op}
\begin{split}
	A(\al',\phi)&=\sqrt{\frac{(2\pi)^3}{V}}\phi_{s'}(\spp')-i\sqrt{\frac{(2\pi)^3}{V}}\sum_s\int
    \frac{d\spp}{(2\pi)^3}\frac{m}{\sqrt{p_0p'_0}}\bar u^{s'}(\spp')M(p',p)u^s(\spp) \phi_s(\spp),
\end{split}
\end{equation}
where the Fourier transform of the mass operator has been introduced
\begin{equation}
	M(p',p):=\int dx dye^{ip'x}M(x,y)e^{-ipy}.
\end{equation}
The two contributions to the amplitude \eqref{amplitude_mass_op} correspond to the Feynman diagrams shown in Fig. \ref{ris:FeynDiag2}.

The probability to record the electron in the state with momenta $(\spp',\spp'+d\spp')$ and the spin vector $\bs\zeta$ is given by
\begin{equation}
\begin{split}
    dP(\bs\zeta,\spp')=\,&\sum_{s_1',s_2'}D_{s_2's_1'}(\bs\zeta)\Big\{\phi_{s_1'}(\spp')\phi^*_{s_2'}(\spp') -i\sum_s\int\frac{d\spp}{(2\pi)^3} \frac{m}{\sqrt{p_0p_0'}} \phi^*_{s_2'}(\spp')\bar{u}^{s_1'}(\spp') M(p',p) u^s(\spp)\phi_s(\spp)+\\
    &+i\sum_s\int\frac{d\spp}{(2\pi)^3} \frac{m}{\sqrt{p_0p_0'}} \phi_{s_1'}(\spp') [\bar{u}^{s_2'}(\spp') M(p',p)u^s(\spp)]^*\phi^*_s(\spp) \Big\}d\spp'+\cdots,
\end{split}
\end{equation}
where the initial state of the electron is described by the wave function $\phi_s(\spp)$ and the terms quadratic in $M$ are omitted. Comparing this expression with \eqref{inclusive_prob_defn}, \eqref{w_defn}, and \eqref{w_contribs}, we reveal that the mass operator of an electron interacting electromagnetically with a hadron on the electron mass-shell reads
\begin{equation}\label{mass_oper}
   M(p',p)=-2\pi e^2 M \ga^\mu \sum_{r,r'}\int d\spk d\spk'\frac{\de(p'+k'-p-k)}{q^2+i0} \frac{\rho^n_{rr'}(\spk,\spk')}{\sqrt{k_0k_0'}} \bar{u}^{r'}(\spk')\Ga_\mu u^r(\spk).
\end{equation}
Notice that this mass operator depends only on $q=p-p'$ and, consequently, it is the Fourier transform of some local operator in the $x$-space. Moreover, the mass operator \eqref{mass_oper} has the form
\begin{equation}\label{mass_oper_2}
    M(q)=e\ga^\mu A^{eff}_\mu(q),
\end{equation}
i.e., this mass operator describes the interaction with some effective electromagnetic field. The corresponding effective current is
\begin{equation}
    j^{eff}_\mu(q):=-q^2 A_\mu^{eff}(q)=2\pi e M \sum_{r,r'}\int d\spk d\spk' \de(k'-k-q) \frac{\rho^n_{rr'}(\spk,\spk')}{\sqrt{k_0k_0'}} \bar{u}^{r'}(\spk')\Ga_\mu u^r(\spk),
\end{equation}
or in the coordinate space
\begin{equation}\label{j_mu_eff_x}
    j^{eff}_\mu(x):=\int \frac{dq}{(2\pi)^4}e^{iqx} j^{eff}_\mu(q)=eM\int\frac{d\spk d\spk'}{(2\pi)^3}  \frac{\rho^n_{rr'}(\spk,\spk')}{\sqrt{k_0k_0'}} e^{i(k'-k)x}\bar{u}^{r'}(\spk')\Ga_\mu u^r(\spk).
\end{equation}
This expression is exactly the average value of the hadron electromagnetic current in the state $\rho_n(\spk,\spk')$. In the case of a pure state \eqref{pure_states}, the average value \eqref{j_mu_eff_x} is reduced to a diagonal matrix element of the hadron current. It is obvious that $\partial^\mu j^{eff}_\mu=0$.

Thus we see that the interference contributions to the inclusive probability of order $\al$ studied in Sec. \ref{Inclus_Probab_Sec} correspond to scattering of an electron by a classical electromagnetic field created by the effective current \eqref{j_mu_eff_x}. The quantum recoil due to scattering of an electron by a hadron is absent in this case \cite{KazSol22}, i.e., these contributions to the inclusive probability are determined by the diagonal matrix element of the hadron electromagnetic current. In a certain sense, the terms of order $\al$ in the inclusive probability to record an electron are the same as if the electron is scattered by a gas of hadrons whose density is given by the modulus squared of the wave function of hadron center-of-mass in the coordinate representation even in the case of scattering by a single hadron. The structure of the electromagnetic current density caused by a nontrivial form of the hadron center-of-mass wave function is crucial for these contributions. For example, the presence of a periodic structure in the hadron wave function results in the Bragg resonances in these contributions, while a large projection of the orbital angular momentum leads to a large effective magnetic moment of the hadron wave packet \cite{IvKarl13prl,IvKarl13pra,KonPotPol,BliokhVErev,LBThY}. In other words, these contributions describe coherent scattering of an electron by a hadron wave packet.

Using expressions \eqref{dens_matrices}, \eqref{hadron_curr_matr_elem}, the density of the hadron electromagnetic current can be written as
\begin{equation}\label{j_eff_expl}
    j^{eff}_\mu(x)=\frac{eM}{2}\int\frac{d\spk d\spk'}{(2\pi)^3} \frac{\rho_n(\spk,\spk')}{\sqrt{k_0k_0'}} e^{i(k'-k)x} \big[\tilde{G}_\mu(\spk,\spk') -\xi^a_n(\spk,\spk')\tau^i_a \tilde{Z}_{i\mu}(\spk,\spk')\big].
\end{equation}
The explicit expressions for $\tilde{G}_\mu$ and $\tilde{Z}_{i\mu}$ are presented in Appendix \ref{App_Traces_Contract}. Let us consider the particular cases (a) and (b) as in the previous section.

In the case (a) for small $q^\mu$, we have
\begin{equation}\label{tild_G}
    \tilde{G}^\mu=\frac{2k_c^\mu}{M}F_e+\cdots,
\end{equation}
in the leading order. The approximate expression for $\tilde{Z}_{i\mu}$ is given in formula \eqref{D_G_Z_approx}. In the nonrelativistic limit,
\begin{equation}\label{tild_Z}
\begin{split}
    \tilde{Z}^{i0}(\spk,\spk')&=-\frac{i}{2M^2}(F_e-2F_m) \e^{ijk} q_jk^c_k+\cdots,\\
    \tilde{Z}^{il}(\spk,\spk')&=\frac{i}{M}\big(\e^{ilj} q_j F_m -\frac{k_c^l}{2M^2} \e^{ijk}q_jk^c_k F_e \big) +\cdots\approx \frac{i}{M}\e^{ilj} q_j F_m,
\end{split}
\end{equation}
where it is assumed that $|F_m|$ is not small in comparison with $|F_e|$ in the last approximate equality. The contributions \eqref{tild_G}, \eqref{tild_Z} possess a clear physical interpretation: the contribution to the effective current proportional to  $\tilde{G}_\mu$ describes the current produced by the charge density; the contribution to the effective current proportional to $\tilde{Z}_{i\mu}$ describes the current associated with the distribution of the particle magnetic moment caused by the particle spin. If the hadron state is not spin polarized, then only the first term proportional to \eqref{tild_G} remains. As is well-known, the current density allows one to recover the wave function of a particle in the nonrelativistic approximation up to a constant phase. Therefore, in this case, the interference terms in the inclusive probability to record an electron can be employed for tomography of the hadron quantum state.

In the case (b), where $F_e\sim q^2/M^2$, the main contribution to the effective current for the spin polarized neutron state comes from the second term in \eqref{j_eff_expl} proportional to $\tilde{Z}_{i\mu}$, i.e., the effective current characterizes a certain distribution of the magnetic moment density of the wave packet of a polarized neutron. If the neutron is in an unpolarized state, then the leading contribution stems from the first term in \eqref{j_eff_expl}. In that case,
\begin{equation}\label{neutron_unpolar}
\begin{split}
    \tilde{G}^0&=2F_e -\frac{q_0^2}{4M^2}F_m+\cdots\approx 2q^2F'_e(0) -\frac{q_0^2}{4M^2}F_m(0),\\
    \tilde{G}^i&=\frac{2k_c^i}{M}\big(F_e -\frac{q^2}{8M^2} F_m\big) -\frac{q_0q^i}{4M^2}F_m +\cdots\approx \frac{2k_c^i}{M}\big[ q^2F'_e(0) -\frac{q^2}{8M^2} F_m(0)\big] -\frac{q_0q^i}{4M^2}F_m(0),
\end{split}
\end{equation}
in the nonrelativistic limit. As we see, the effective electromagnetic current is small, inasmuch as it is proportional to $\spq^2/M^2$, but different from zero.

The mass operator \eqref{mass_oper_2} enters into the effective Dirac equation describing an electron in the electromagnetic field produced by the hadron wave packet. It is anticipated that the bound states of this equation give rise to the resonances in the inclusive probability to record an electron scattered by a hadron wave packet. In order to analyze such states, one needs, at first, to generalize the expression for $M(q)$ obtained above to the case where the external electron lines are out of the mass-shell. Nevertheless, for small virtualities of the external lines, $|p^2/m^2-1|\ll1$, $|p'^2/m^2-1|\ll 1$, one may exploit expression \eqref{mass_oper} derived on the mass-shell. The formal definition of the mass operator and the other one-particle irreducible contributions to the Green functions is convenient to provide in the framework of the $in$-$in$ formalism developed for evaluation of probabilities and averages of observables \cite{MartSchw,Keld64,CSHY85,DeWGAQFT.11}. In this case, the modification of the standard approach consists in that one should define the $in$-$in$ effective action in the presence of a hadron with given wave function in the $in$ state. A detailed analysis of the respective contributions to the effective action will by given elsewhere.

It is not difficult to see that the bound states appearing in solving the Dirac equation with the mass operator \eqref{mass_oper_2} do not coincide with the bound states arising due to interaction of the electron with the hadron, where the wave function of the hadron is concentrated at some point in the $x$-space as in the case of hydrogenic bound states. In order to obtain the ``standard'' bound states, one has to sum an infinite number of Feynman diagrams, for example, by solving the Bethe-Salpeter equation or by introducing the corresponding bound states directly to the initial Lagrangian.

\section{Conclusion}

Let us sum up the results. We have considered elastic electromagnetic scattering of an electron by a spin one-half hadron. Scattering of an electron by a muon also falls under our study. The initial state of the electron and the hadron has been taken to be a separable mixed state of a general form. It turns out that the leading nontrivial contribution to the inclusive probability to record an electron in such a process is of order $\al$ rather than $\al^2$ as in the Rosenbluth formula \cite{Rosen50,LandLifQED} for the differential cross-section. This contribution appears only for the initial states of the electron and the hadron distinct from plane waves. It stems from the interference of the trivial contribution to the $S$-matrix and the leading contribution to the connected part of the $S$-matrix \cite{KazSol22}.

We have obtained the explicit expression for this interference contribution \eqref{inclusive_prob} and have described its properties. The particular case where the electron and the hadron are prepared in the unpolarized Gaussian states has been also scrutinized in Appendix \ref{App_Gauss_wave_p}. We have established that in interpreting the interference contribution one may think of the electron as scattered by the classical electromagnetic field created by the hadron electromagnetic current averaged with respect to the free evolving density matrix of hadron center-of-mass \eqref{j_mu_eff_x}. In particular, there is no quantum recoil experienced by a hadron due to scattering on an electron in this contribution. Moreover, this interpretation is valid even in the case of scattering by a single hadron. We have found that the interference contribution can be employed to reconstruct the wave function of a scattered hadron with the aid of the inclusive probability to detect the electron. We have derived the effective electromagnetic potential and the corresponding effective current \eqref{j_mu_eff_x} that can substitute for the hadron wave packet in describing the interference term in the inclusive probability. This effective current is precisely the average hadron electromagnetic current. It entails that the interference contribution corresponds to coherent scattering of an electron by a hadron wave packet and so the insights from classical electrodynamics and scattering theory of electrons by classical electromagnetic fields can be employed.

The nontrivial form of the hadron wave packet strongly affects the interference contribution. For example, the periodic structure of the hadron wave packet gives rise to the Bragg resonances in the interference contribution, the symmetries of the hadron wave packet lead to the respective selection rules \cite{KazSol22}, the large projection of the angular momentum of a charged hadron causes the large magnetic moment of the effective current \cite{IvKarl13prl,IvKarl13pra,KonPotPol,BliokhVErev,LBThY} with the corresponding imprints on the scattering data, the higher magnetic and electric multipoles of the hadron wave packet \cite{KarlZhev19,PupKarl22} also influence the interference contribution. We have derived the effective electromagnetic current for a neutron in both spin polarized states and unpolarized ones. In the latter case, the effective current is small but nonzero and is given by \eqref{j_eff_expl}, \eqref{neutron_unpolar}.

The effective electromagnetic potential enters into the mass operator of the electron. The Dirac equation with this mass operator possesses bound states. It is expected that these bound states reveal itself as resonances in the interference contribution in addition to the standard resonances coming from the bound states of an electron and a hadron. We leave a detailed investigation of the additional bound states and a development of the corresponding formalism for a future research. Another possible direction of research related to the results of the present paper is the study of impact of a measurement on the interference contribution to the inclusive probability. As is postulated in quantum mechanics, a measurement results in a collapse of the wave function of a particle. Since the interference term depends on the form of the wave packet and is determined by the electromagnetic current constructed in terms of this wave function, it is anticipated that the measurement affects severely this contribution to the inclusive probability.

\paragraph{Acknowledgments.}

We appreciate the anonymous referee for valuable comments. This study was supported by the Tomsk State University Development Programme (Priority-2030).

\appendix
\section{Traces and contractions}\label{App_Traces_Contract}

In evaluating the inclusive probability and the effective current there appear the following traces of products of $\ga$ matrices:
\begin{equation}\label{funcdef}
\begin{split}
	D(\spp,\spp')&:=\frac{\Sp\left[(m+\hat p)\frac{\ga^0+1}{2}(m+
    \hat p')\right]}{2m\sqrt{p_0+m}\sqrt{p'_0+m}} = \frac{4m^2 +4m p^0_c-q^2}{2m\sqrt{(p^0_c+m)^2-q_0^2/4}},\\
	G^\mu(\spp,\spp')&:=\frac{\Sp\left[\ga^\mu (m+\hat p)\frac{\ga^0+1}{2}(m+
    \hat p')\right]}{2m\sqrt{p_0+m}\sqrt{p'_0+m}}=\frac{4p_c^\mu(p_c^0+m)+\eta^{\mu0}q^2-q_0q^\mu}{2m\sqrt{(p^0_c+m)^2-q_0^2/4}},\\
	Z^{i\mu}(\spp,\spp')&:=-\frac{\Sp\left[\ga^\mu (m+\hat p)\ga^i \ga^5\frac{\ga^0+1}{2}(m+
    \hat p')\right]}{2m\sqrt{p_0+m}\sqrt{p'_0+m}}=i\frac{\e^{i\mu\nu\rho}q_\nu (p^c_\rho+m\de^0_\rho) }{m\sqrt{(p^0_c+m)^2-q_0^2/4}},\\
	X^i(\spp,\spp')&:=-\frac{\Sp\left[(m+\hat p)\ga^i \ga^5\frac{\ga^0+1}{2}(m+
    \hat p')\right]}{2m\sqrt{p_0+m}\sqrt{p'_0+m}}=-i\frac{\e^{i\nu\rho0} q_\nu p^c_\rho}{m\sqrt{(p^0_c+m)^2-q_0^2/4}},
\end{split}
\end{equation}
where $p_c:=(p+p')/2$, $\e^{\mu\nu\rho\s}$ is the Levi-Civita symbol, and $\e^{0123}=1$. It is clear that
\begin{equation}
    q_\mu p_c^\mu=q_\mu k_c^\mu=0,\qquad k_c^2=M^2-q^2/4,\qquad p_c^2=m^2-q^2/4.
\end{equation}
There are the relations
\begin{equation}\label{rels}
\begin{gathered}
    D(\spp,\spp')=\frac{p^c_\mu}{m}G^\mu(\spp,\spp'),\qquad X^i(\spp,\spp')=-Z^{i0}(\spp,\spp'),\qquad q_\mu G^\mu(\spp,\spp')=q_\mu Z^{i\mu}(\spp,\spp')=0,\\
    [G^\mu(\spp,\spp')]^*=G^\mu(\spp,\spp'),\qquad[Z^{i\mu}(\spp,\spp')]^*=-Z^{i\mu}(\spp,\spp').
\end{gathered}
\end{equation}
To describe the contributions of the hadron electromagnetic current, it is useful to introduce the notation
\begin{equation}
\begin{split}
    \tilde{G}^\mu(\spk,\spk')&=G^\mu(\spk,\spk')F_m(q^2)+\frac{Mk_c^\mu}{k_c^2} D(\spk,\spk') [F_e(q^2)-F_m(q^2)],\\
    \tilde{Z}^{i\mu}(\spk,\spk')&=Z^{i\mu}(\spk,\spk')F_m(q^2)+\frac{Mk_c^\mu}{k_c^2} X^i(\spk,\spk') [F_e(q^2)-F_m(q^2)],
\end{split}
\end{equation}
where $G^\mu(\spk,\spk')$, $D(\spk,\spk')$, $Z^{i\mu}(\spk,\spk')$, and $X^i(\spk,\spk')$ are defined as in \eqref{funcdef} but with the replacements $m\rightarrow M$ and $q\rightarrow-q$. Employing the relations \eqref{rels}, we can rewrite $\tilde{G}^\mu$ and $\tilde{Z}^{i\mu}$ as
\begin{equation}
\begin{split}
    \tilde{G}^\mu(\spk,\spk')&=G^\mu(\spk,\spk')F_m(q^2)+\frac{k_c^\mu k_c^\nu}{k_c^2} G_\nu(\spk,\spk') [F_e(q^2)-F_m(q^2)],\\
    \tilde{Z}^{i\mu}(\spk,\spk')&=Z^{i\mu}(\spk,\spk')F_m(q^2)-\frac{Mk_c^\mu}{k_c^2} Z^{i0}(\spk,\spk') [F_e(q^2)-F_m(q^2)].
\end{split}
\end{equation}
The following relations hold
\begin{equation}\label{tilde_G_prop}
    q_\mu \tilde{G}^\mu(\spp,\spp')=q_\mu \tilde{Z}^{i\mu}(\spp,\spp')=0,\qquad [\tilde{G}^\mu(\spp,\spp')]^*=\tilde{G}^\mu(\spp,\spp'), \qquad [\tilde{Z}^{i\mu}(\spp,\spp')]^*=-\tilde{Z}^{i\mu}(\spp,\spp').
\end{equation}
Then
\begin{equation}\label{contractions}
\begin{split}
    \tilde{G}^\mu_nG^e_\mu &=F_m G^\mu_n G_\mu^e +(F_e-F_m)\frac{k^c_\mu G^\mu_e k^c_\nu G^\nu_n}{k_c^2},\\
    \tilde{G}^\mu_n Z^i_{e\mu}&=F_m G^\mu_n Z^i_{e\mu} +(F_e-F_m)\frac{k^c_\mu Z^{i\mu}_e k^c_\nu G^\nu_n}{k_c^2},\\
    \tilde{Z}^i_{n\mu}G_e^\mu &=F_m Z^i_{n\mu} G^\mu_e -(F_e-F_m)\frac{M}{k_c^2}Z^{i0}_n  k^c_\mu G^\mu_e,\\
    \tilde{Z}^{i\mu}_{n} Z^e_{j\mu} &= F_mZ^{i\mu}_{n} Z^e_{j\mu} -(F_e-F_m)\frac{M}{k_c^2}Z^{i0}_n k^\mu_c Z^e_{j\mu},
\end{split}
\end{equation}
where, for brevity, we introduce the notation $G^\mu_n\equiv G^\mu(\spk,\spk')$, $G_e^\mu:=G^\mu_e(\spp,\spp')$, and so on. The  contractions appearing in the inclusive probability have the form
\begin{equation}\label{contractions_1}
\begin{split}
    G^n_\mu G^\mu_e&=\frac{16(k_cp_c)(k^0_c+M)(p^0_c+m) +4q^2[k^0_c(k^0_c+M) +p^0_c(p^0_c+m)] +q^4-q_0^2q^2 }{4mM\sqrt{[(p^0_c+m)^2-q_0^2/4][(k^0_c+M)^2-q_0^2/4]}},\\
    G^n_\mu Z^{i\mu}_e&=i\frac{4(k^0_c+M)\e^{i\mu\nu\rho}k^c_\mu q_\nu(p^c_\rho+m\de^0_\rho)-q^2\e^{0i\nu\rho}q_\nu p^c_\rho }{2mM\sqrt{[(p^0_c+m)^2-q_0^2/4][(k^0_c+M)^2-q_0^2/4]}},\\
    G^e_\mu Z^{i\mu}_n&=-i\frac{4(p^0_c+m)\e^{i\mu\nu\rho}p^c_\mu q_\nu(k^c_\rho+M\de^0_\rho)-q^2\e^{0i\nu\rho}q_\nu k^c_\rho }{2mM\sqrt{[(p^0_c+m)^2-q_0^2/4][(k^0_c+M)^2-q_0^2/4]}},\\
    Z^{i\mu}_{n} Z^e_{j\mu}&= \frac{\e^{i\mu\nu\rho}\e_{j\mu\nu'\rho'}q_\nu (k^c_\rho+M\de^0_\rho) q^{\nu'} (p_c^{\rho'}+m\de_0^{\rho'}) }{mM\sqrt{[(p^0_c+m)^2-q_0^2/4][(k^0_c+M)^2-q_0^2/4]}},
\end{split}
\end{equation}
where $\e_{0123}=-1$. Besides,
\begin{equation}
\begin{split}
    k^c_\mu G^\mu_e &=\frac{4(k_cp_c)(p^0_c+m)+k^0_cq^2}{2m\sqrt{(p^0_c+m)^2-q_0^2/4}},\\
    k^c_\mu Z^{i\mu}_e &=i\frac{\e^{i\mu\nu\rho}k^c_\mu q_\nu (p^c_\rho+m\de^0_\rho)}{m\sqrt{(p^0_c+m)^2-q_0^2/4}}.
\end{split}
\end{equation}
One can get rid of the contraction of the two Levi-Civita symbols entering into \eqref{contractions_1}. However, this does not lead to a considerable simplification of the expression.

Let us find the approximate expressions for the contractions obtained above. Suppose that
\begin{equation}\label{small_recoil}
    \spq^2/(k^0_c)^2\ll1,\qquad \spq^2/(p^0_c)^2\ll1.
\end{equation}
Recall that $\spq^2$ is of order of the momentum dispersion in the wave packet. Therefore, the above estimates are fulfilled as a rule. In that case, keeping only the leading contributions in $q^\mu$, we come to
\begin{equation}\label{D_G_Z_approx}
\begin{split}
    D(\spp,\spp')=\,& 2+\frac{q_0^2}{4(p^0_c+m)^2}-\frac{q^2}{2m(p^0_c+m)}+\cdots,\\
	G^\mu(\spp,\spp')=\,&\frac{2p_c^\mu}{m}+\frac{q^2\eta^{0\mu}-q_0q^\mu}{2m(p^0_c+m)}+\frac{q_0^2 p_c^\mu}{4m(p_c^0+m)^2}+\cdots,\\
	Z^{i\mu}(\spp,\spp')=\,&i\frac{\e^{i\mu\nu\rho}q_\nu (p^c_\rho+m\de^0_\rho) }{m (p^0_c+m)}+\cdots,\\
    \tilde{G}^\mu(\spk,\spk')=\,&\frac{2k_c^\mu}{M}\Big(1+\frac{q_0^2}{8(k_c^0+M)^2}+\frac{q^2k_c^0}{4M^2(k_c^0+M)}\Big)F_e +\Big(\frac{q^2\eta^{\mu0}-q_0q^\mu}{2M(k_c^0+M)} -\frac{q^2k_c^0k_c^\mu}{2M^3(k_c^0+M)} \Big)F_m+\cdots,\\
    \tilde{Z}^{i\mu}(\spk,\spk')=\,&\frac{-i}{M(k_c^0+M)} \Big[\frac{k_c^\mu}{M}\e^{0i\nu\rho}q_\nu k^c_\rho(F_e-F_m) +\e^{i\mu\nu\rho}q_\nu (k^c_\rho+M\de^0_\rho)F_m \Big]+\cdots.
\end{split}
\end{equation}
In the leading order in $q^\mu$, the contractions \eqref{contractions} become
\begin{equation}
\begin{split}
    \tilde{G}^\mu_nG^e_\mu &=\frac{4(k_cp_c)}{mM}F_e+\cdots,\\
    \tilde{G}^\mu_n Z^i_{e\mu}&=\frac{2iF_e}{mM(p_c^0+m)}\e^{i\mu\nu\rho}k^c_\mu q_\nu(p^c_\rho+m\de^0_\rho)+\cdots,\\
    \tilde{Z}^i_{n\mu}G_e^\mu &= \frac{-2i}{mM(k_c^0+M)}\Big[\e^{i\mu\nu\rho} p^c_\mu q_\nu (k^c_\rho+M\de^0_\rho)F_m +\frac{(k_cp_c)}{M}\e^{0i\nu\rho}q_\nu k^c_\rho (F_e-F_m)\Big]+\cdots,\\
    \tilde{Z}^{i\mu}_{n}Z^e_{j\mu} &=\frac{\e^{i\mu\nu\rho}\e_{j\mu\nu'\rho'}q_\nu (k^c_\rho+M\de^0_\rho) q^{\nu'} (p_c^{\rho'}+m\de_0^{\rho'}) }{mM (p^0_c+m)(k^0_c+M)}F_m +\frac{\e^{0i\nu\rho}q_\nu k^c_\rho\e_{j\mu'\nu'\rho'} k_c^{\mu'}q^{\nu'}(p_c^{\rho'}+m\de^{\rho'}_0)}{mM^2 (p^0_c+m)(k^0_c+M)}  (F_e-F_m)+\cdots.
\end{split}
\end{equation}
In the case of a nonrelativistic hadron, one can put $k_c^0\approx M\gg|\spk_c|$. In the ultrarelativistic limit for an electron, we have $p_c^0\approx|\spp_c|\gg m$.

\section{Gaussian wave packets}\label{App_Gauss_wave_p}

In this section, we consider the particular case of the general formula \eqref{w23_nonrel} for the inclusive probability to record an electron scattered by a charged hadron in the case of Gaussian initial wave functions of the particles. We obtain the explicit expressions for the interference contribution and for the term $w^{(4)}_{s_1's_2'}$.

We assume that the initial density matrices of the electron and the hadron center-of-mass take the form
\begin{equation}
\begin{split}
    \rho^e_{ss'}(\spp,\spp')&=\frac{\de_{ss'}}{2(2\pi)^{3/2}\s^3} \exp\Big[-\frac{(\spp-\spp_0)^2}{4\s^2} -\frac{(\spp'-\spp_0)^2}{4\s^2} -i\mathbf{g}_e(\spp-\spp')\Big],\\
    \rho^n_{rr'}(\spk,\spk')&=\frac{\de_{rr'}}{2(2\pi)^{3/2}\s_n^3} \exp\Big[-\frac{\spk^2}{4\s^2_n} -\frac{\spk'^2}{4\s^2_n} -i\mathbf{g}_n(\spk-\spk')\Big],\\
\end{split}
\end{equation}
where $\s^2$ is the dispersion of momenta in the electron wave packet, $\s^2_n$ is the dispersion of momenta in the hadron center-of-mass wave packet, $\spp_0$ is the average momentum in the electron wave packet, $\mathbf{g}_e$ and $\mathbf{g}_n$ are the average positions of the electron and the hadron in the coordinate space. For simplicity, the initial states of the electron and the hadron are supposed to be unpolarized.

Let us assume that $|\bs\be'_n|\ll |\bs\be'_e|$ and the transferred momentum, $|\spq_\perp|\lesssim\s$, is much less than the typical scale of variations of the form factor $F_e(-\spq_\perp^2)$, then we have
\begin{equation}
    F_e(-\spq_\perp^2)\approx-Z.
\end{equation}
Under these assumptions, it follows from formula \eqref{w23_nonrel} that
\begin{equation}
    w^{(2)}_{s'_1s'_2}+w^{(3)}_{s'_1s'_2}=-\frac{Ze^2\delta_{s_1's_2'}}{|\bs{\be}'_e|(2\pi)^{7/2}\sigma^3} e^{-\frac{\Delta\spp^2}{2\sigma^2}}\int \frac{d\spq_\perp}{\spq_\perp^2}\im\exp\Big[- \frac{\spq_\perp^2}{4\Si^2} +\frac{(\spp_{0\perp},\spq_\perp)}{2\sigma^2} -i\spq_\perp \spg_\perp\Big],
\end{equation}
where $\Delta\spp:=\spp'-\spp_0$, $\Si^{-2}=\s^{-2}+\s_n^{-2}/2$, $\spg:=\spg_e-\spg_n$ is the impact parameter, and we have taken into account that $(\spp',\spq_\perp)=0$. Using the representation,
\begin{equation}
	1/\spq_\perp^2=\int_0^\infty d\al e^{-\al \spq_\perp^2},
\end{equation}
and performing the Gaussian integration, we come to
\begin{equation}
    \int \frac{d\spq_\perp}{\spq_\perp^2}\im\exp\Big[- \frac{\spq_\perp^2}{4\Si^2} +\frac{(\spp_{0\perp},\spq_\perp)}{2\sigma^2} -i\spq_\perp \spg_\perp\Big]=\pi\int_0^1\frac{d\al}{\al}\im e^{-\al\Si^2 v_+},
\end{equation}
where $v_\pm:=(\spg_\perp\pm i\spp_{0\perp}/(2\s^2))^2$. Thus,
\begin{equation}
    w^{(2)}_{s'_1s'_2}+w^{(3)}_{s'_1s'_2}=w^{(1)}_{s_1's_2'}\frac{2Z\al}{|\bs{\be}'_e|}\im h(\Si^2 v_+).
\end{equation}
For brevity, we have introduced the notation
\begin{equation}
    h(z):=\ga+\ln z+\Ga(0,z),
\end{equation}
where $\ga$ is the Euler constant. The function $h(z)$ is an entire real analytic function of $z$ and $h(0)=0$.

\begin{figure}[tp]
\centering
\includegraphics*[width=0.3\linewidth]{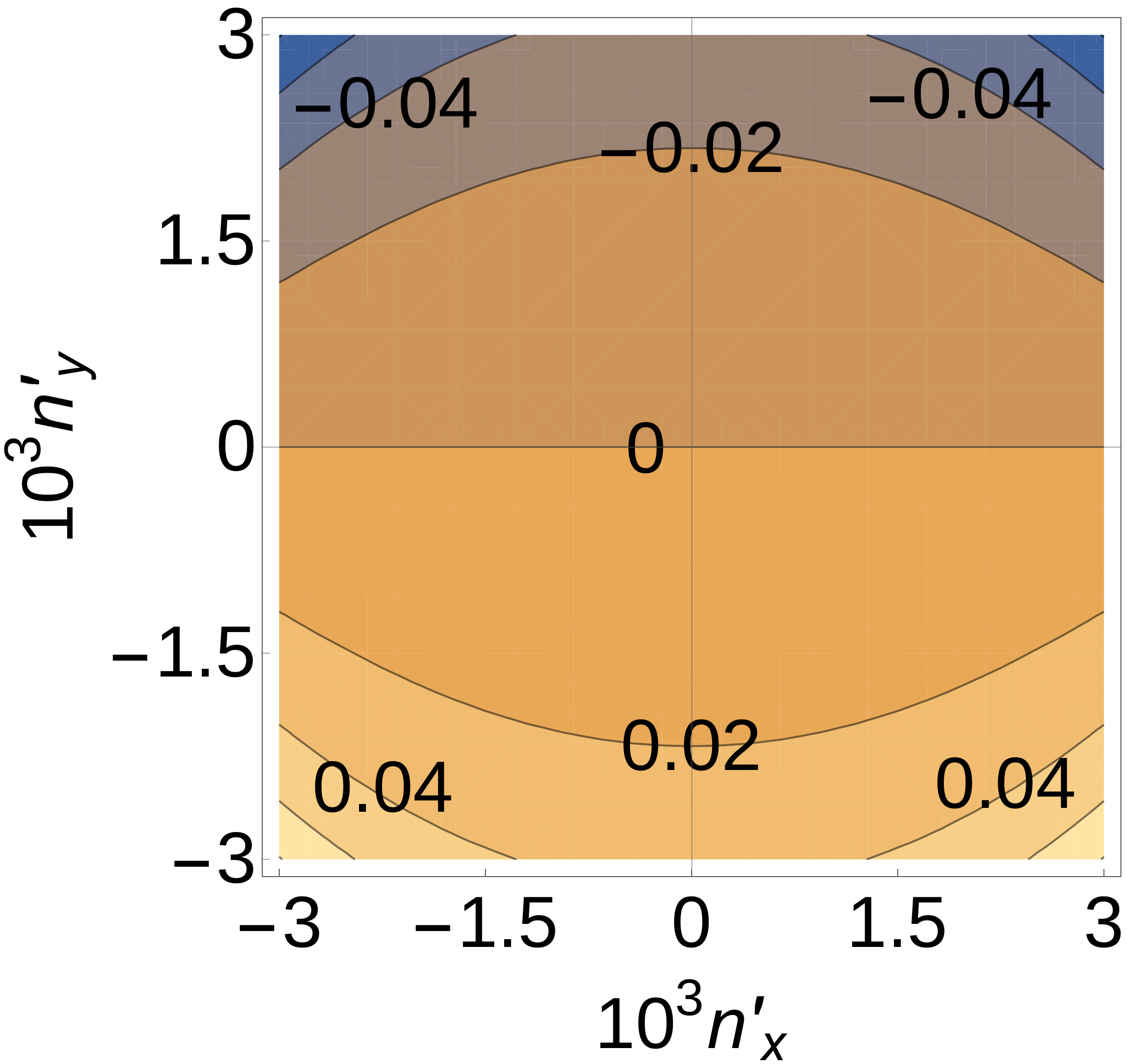}\;
\includegraphics*[width=0.3\linewidth]{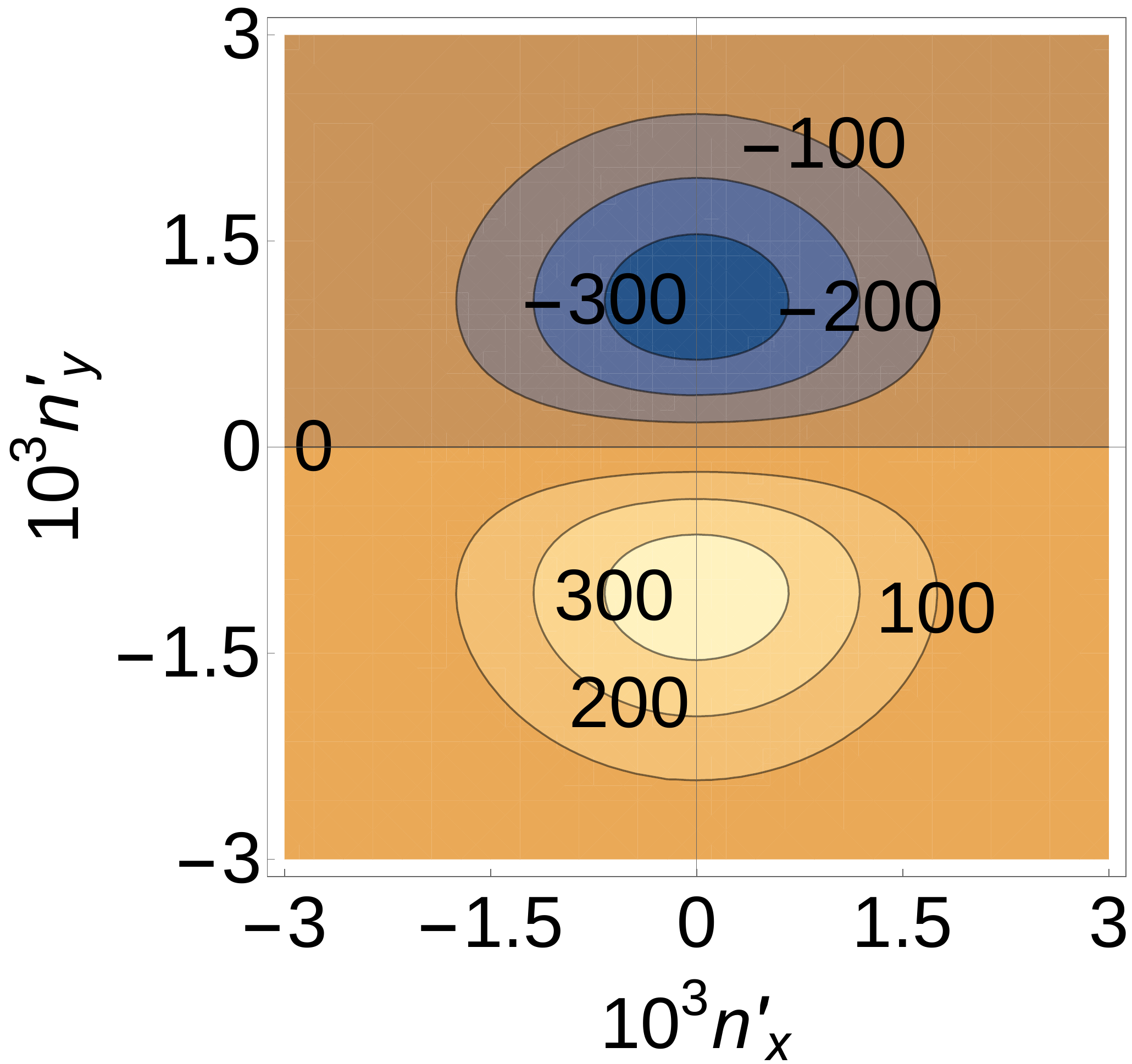}\;
\includegraphics*[width=0.3\linewidth]{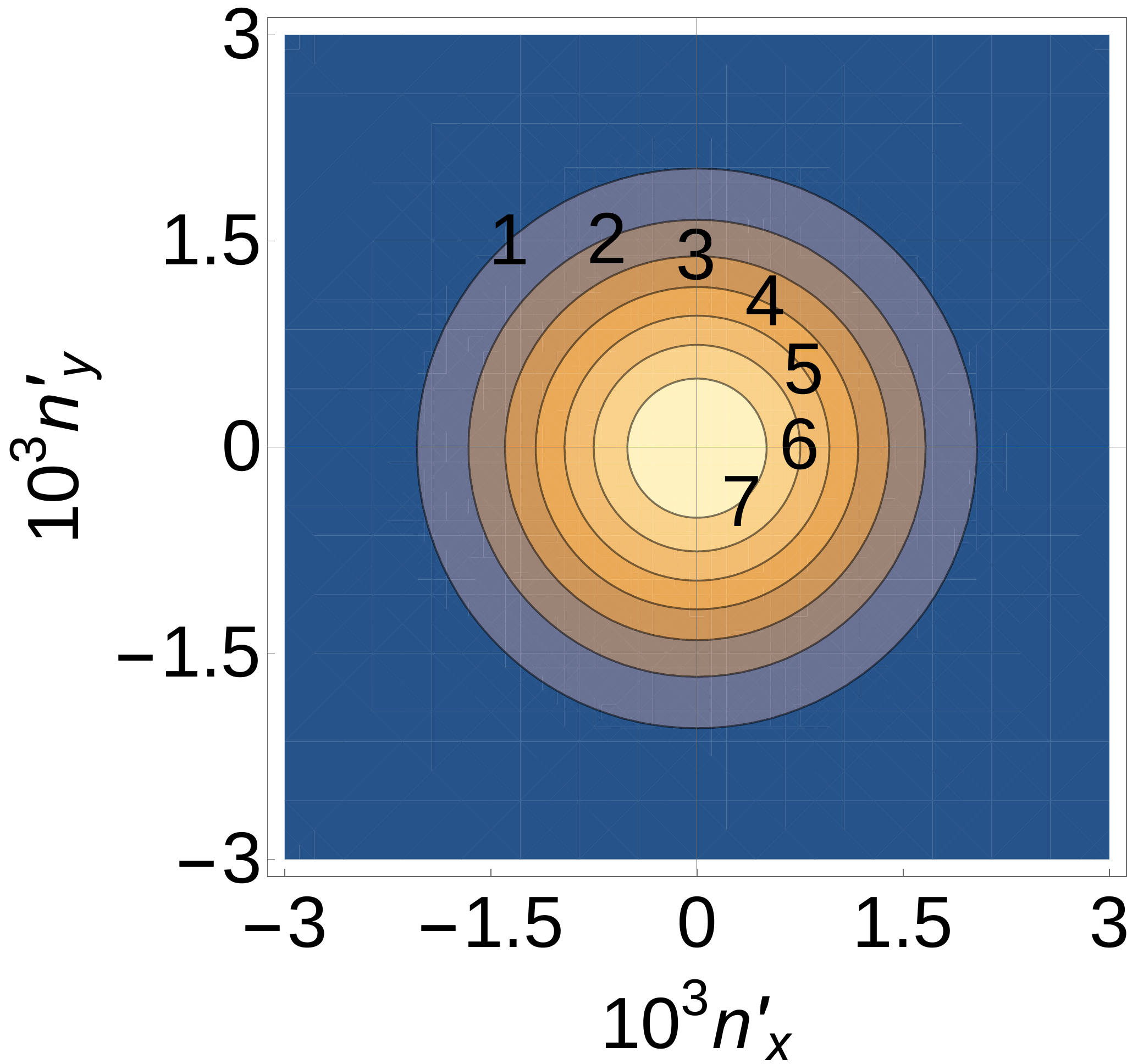}
\caption{{\footnotesize The effect of the interference contribution on the inclusive probability as follows from formula \eqref{inclus_probab_Gaussian}. The average electron momentum $p_0=10$ MeV, the normal deviation of momenta in the electron and hadron wave packets $\s=\s_n=10$ keV, the impact parameter $\spg=(0,1/\s,0)$, the $z$ axis is directed along $\spp_0$, and $Z=1$. Left panel: The ratio of the interference contribution to the main contribution. Middle panel: The separate contribution of the interference term to the electron density matrix after scattering. Right panel: The density matrix of the escaped electron \eqref{inclus_probab_Gaussian} divided by $10^4$.}}
\label{interference_plots}
\end{figure}

The variable $v_+$ does not depend on the modulus of momentum of the recorded electron. Therefore, one can readily integrate the electron density matrix after scattering \eqref{w_defn} with respect to it. Since
\begin{equation}\label{Gauss_int_rad}
	\int_0^\infty d|\spp'| |\spp'|^2 e^{-\frac{(\spp'-\spp_0)^2}{2\sigma^2}}=2\sigma^3
    e^{-\frac{\spp_0^2}{2\sigma^2}+\frac{(\spp_0,\spn')^2}{4\sigma^2}}D_{-3}\Big(-\frac{\spp_0\spn'}{\sigma}\Big),
\end{equation}
where $\spn':=\spp'/|\spp'|$ and $D_\nu(z)$ is the parabolic cylinder function \cite{GrRy}, we eventually arrive at
\begin{equation}\label{inclus_probab_Gaussian}
	\int_0^\infty d|\spp'||\spp'|^2
    \big(w^{(1)}_{s_1's_2'}+w^{(2)}_{s_1's_2'}+w^{(3)}_{s_1's_2'}\big)=\frac{\de_{s'_1s'_2}}{(2\pi)^{3/2}} e^{-\frac{\spp_0^2}{2\sigma^2}+\frac{(\spp_0\spn')^2}{4\sigma^2}}D_{-3}\Big(-\frac{\spp_0\spn'}{\sigma}\Big) \Big[1+\frac{2Z\al}{|\bs{\be}'_e|}\im h(\Si^2 v_+)\Big].
\end{equation}
Substituting this into \eqref{inclusive_prob_gen}, we obtain the probability to detect the electron escaping along the given direction $\spn'$ up to the second order in the coupling constant $e$. The plots of the contributions to \eqref{inclus_probab_Gaussian} are given in Fig. \ref{interference_plots}.

Let us find the explicit expression for the contribution $w^{(4)}_{s_1's_2'}$ in the domain of parameters as in the considerations above. Notice that apart from this contribution there are the loop corrections to the inclusive probability of order $e^4$. For consistency, they should also be taken into account, but we leave their investigation for a future research. The contribution $w^{(4)}_{s_1's_2'}$ contains the sums
\begin{equation}
\begin{split}
	\sum_{r,\tilde r,r'}\rho^n_{r \tilde r}\bar u^{r'}(\spk')\Ga_\mu u^r(\spk) \bar u^{\tilde r}(\tilde \spk)
    \Ga_\nu u^{r'}(\spk')= &\,\frac{\rho_n}{4} \Big[\tilde G_\mu \tilde{G}'_\nu -\xi_a^n \tau_{ai} (\tilde{G}_\mu \tilde{Z}'^i_\nu +\tilde{Z}^i_\mu \tilde{G}'_\nu) -\tilde{Z}^i_\mu \tilde{Z}'_{i\nu}-\\
    &-i \xi_a^n\tau^k_a\e_{kij} \tilde{Z}^i_\mu \tilde{Z}'^j_\nu\Big],\\
	\sum_{s,\tilde s}\rho^e_{s \tilde s}\bar u^{s'_1}(\spp')\ga_\mu u^s(\spp)\bar u^{\tilde s}
	(\tilde \spp)\ga_\nu u^{s'_2}(\spp')=&\,\frac{\rho_e}{8}\Big[ \delta_{s_1' s_2'} \Big(G_\mu G'_\nu -\xi_a^e \tau_{ai}
    (G_\mu Z'^i_\nu +Z^i_\mu G'_\nu) -Z^i_\mu Z'_{i\nu}-\\
    &-i \xi_a^e\tau^k_a\e_{kij} Z^i_\mu Z'^j_\nu\Big)
    + (\sigma_a)_{s_1's_2'} \Big(\xi^e_a G_\mu G'_\nu-\\
    &-\tau_{ai} (Z^{i}_\mu G'_{\nu}+G_{\mu} Z'^{i}_\nu)+\\
    &+i\e_{abc} \xi^e_b \tau_{ci}(Z^{i}_\mu G'_{\nu} -G_{\mu} Z'^{i}_\nu) +\\
    &+\xi^e_a Z^{i}_\mu Z'_{i\nu} +(i\tau^k_a\e_{kij} +\xi^e_b\tau_{(bi}\tau_{a)j}) Z^{i}_\mu Z'^{j}_\nu\Big) \Big],
\end{split}
\end{equation}
where the parentheses at a pair of indices mean symmetrization without the factor $1/2$ and
\begin{equation}
\begin{aligned}
    G_\mu&\equiv G_\mu(\spp,\spp'),&\qquad G'_\nu&\equiv G_\nu(\spp',\tilde \spp),&\qquad Z_\mu&\equiv Z_\mu(\spp,\spp'),&\qquad Z'_\nu&\equiv Z_\nu(\spp',\tilde \spp),\\
    \tilde{G}_\mu&\equiv \tilde{G}_\mu(\spk,\spk'),&\qquad \tilde{G}'_\nu&\equiv \tilde{G}_\nu(\spk',\tilde \spk),&\qquad \tilde{Z}_\mu&\equiv \tilde{Z}_\mu(\spk,\spk'),&\qquad \tilde{Z}'_\nu&\equiv \tilde{Z}_\nu(\spk',\tilde \spk),\\
    &&\rho_n&\equiv\rho_n(\spk,\tilde{\spk}),&\qquad\rho_e&\equiv\rho_e(\spp,\tilde{\spp}).&&
\end{aligned}
\end{equation}
As follows from the the explicit expressions for these function given in Appendix \ref{App_Traces_Contract}, only the terms $G_\mu G'_\nu$ and $\tilde{G}_\mu\tilde{G}'_\nu$ survive for a charged hadron in the leading order in $q_\mu$.

Suppose as above that the initial electron and hadron states are unpolarized, viz., $\xi^e_a=0$ and $\xi^n_a=0$. Then carrying out all the calculations as in Sec. \ref{Inclus_Probab_Sec} and making the same approximations, we obtain
\begin{equation}
    w^{(4)}_{s'_1s'_2}=\frac{Z^2e^4\de_{s'_1s'_2}}{2(2\pi)^4\bs\be'^2_e} \int\frac{d\spq_\perp d\tilde{\spq}_\perp d\spk'}{(\spq^2_\perp-i0) (\tilde{\spq}^2_\perp+i0)}\rho_e(\spp'+\spq_\perp,\spp'+\tilde{\spq}_\perp) \rho_n(\spk'-\spq_\perp,\spk'-\tilde{\spq}_\perp).
\end{equation}
It is not difficult to see that this integral is infrared divergent. As in \cite{WeinbergB.12}, we introduce a small infrared cutoff $\la$ so that the photon momenta $|\spq_\perp|<\la$ and $|\tilde{\spq}_\perp|<\la$ are not realized. Then the $i0$ additions in the denominator can be omitted. The integration over $\spk'$ gives rise to
\begin{equation}
    \int d\spk' \rho_n(\spk'-\spq_\perp,\spk'-\tilde{\spq}_\perp)=\exp\big[-\frac{1}{8\s_n^2}(\spq_\perp-\tilde{\spq}_\perp)^2 +i(\spq_\perp-\tilde{\spq}_\perp)\spg^n_\perp\big].
\end{equation}
The infrared cutoff is introduced by employing the Mellin representation of the step function
\begin{equation}
    \theta(|\spq_\perp|-\la)\theta(|\tilde{\spq}_\perp|-\la)=\int_C\frac{d\nu_1 d\nu_2}{(2\pi i)^2\nu_1\nu_2} \Big(\frac{\spq_\perp^2}{\la^2}\Big)^{\nu_1} \Big(\frac{\tilde{\spq}_\perp^2}{\la^2}\Big)^{\nu_2},
\end{equation}
where the contours $C$ in both the $\nu_1$-plane and the $\nu_2$-plane run upwards parallel to the imaginary axis and slightly from the right of it. Then using the representation,
\begin{equation}
    (\spq_\perp^2)^{\nu_1-1} (\tilde{\spq}_\perp^2)^{\nu_2-1}=\int_0^\infty \frac{d\al_1 d\al_2\al_1^{-\nu_1} \al_2^{-\nu_2}}{\Ga(1-\nu_1)\Ga(1-\nu_2)} e^{-\al_1\spq^2_\perp-\al_2\tilde{\spq}^2_\perp},
\end{equation}
and performing the Gaussian integration with respect to $\spq_\perp$ and $\tilde{\spq}_\perp$, we come after a little algebra to
\begin{equation}\label{w4_int}
\begin{split}
    w^{(4)}_{s'_1s'_2}=&\,\frac{\pi^2Z^2e^4}{(2\pi)^4\bs\be'^2_e} w^{(1)}_{s'_1s'_2}\int_C \frac{d\nu_1 d\nu_2}{(2\pi i)^2\nu_1\nu_2} \frac{(4\Si^2/\la^2)^{\nu_1+\nu_2}}{\Ga(1-\nu_1)\Ga(1-\nu_2)}\times\\
    &\times\int_0^1 dadc a^{\nu_1-1} c^{\nu_2-1} \frac{(1-a)^{-\nu_1}(1-c)^{-\nu_2}}{1-u^2 ac} \exp\Big(-\frac{a \Si^2v_+ +c\Si^2v_- -2yu ac}{1-u^2 ac}\Big),
\end{split}
\end{equation}
where $y:=\Si^2(\spg_\perp^2+\spp_{0\perp}^2/(4\s^4))$ and $u:=\Si^2/(2\s_n^2)$.

In order to obtain the asymptotics of expression \eqref{w4_int} for $\la\rightarrow0$, we ought to shift the integration contours $C$ in the $\nu_1$ and $\nu_2$ planes to the left to the domain where $\re(\nu_1+\nu_2)<0$ taking into account the singularity structure (the poles) of the integrand (see for details \cite{GSh,ParKam01,Wong,KalKaz3}). After that the integral with respect to $\nu_1$ and $\nu_2$ can be discarded as it tends to zero when $\la\rightarrow0$. Denoting concisely the integral on the last line in \eqref{w4_int} as
\begin{equation}
    I:=\int_0^1 dadc a^{\nu_1-1} c^{\nu_2-1} f(a,c),
\end{equation}
we can write
\begin{equation}\label{I_int}
\begin{split}
    I=&\,\int_0^1 dadc a^{\nu_1-1} c^{\nu_2-1} \big[f(a,c) -f(0,c) -f(a,0) +f(0,0)\big]+\\
    &+ \int_0^1 dadc a^{\nu_1-1} c^{\nu_2-1}\big[f(0,c) +f(a,0) -f(0,0)\big].
\end{split}
\end{equation}
We need to find the divergent and finite parts of this expression with respect to $\nu_1$ and $\nu_2$ near the points $\nu_1=0$, $\nu_2=0$. The integral on the first line of \eqref{I_int} is regular at the these points and so we can put $\nu_1=\nu_2=0$ in it. The integral on the second line can easily be expanded near these points with the result
\begin{equation}
    \int_0^1 dadc a^{\nu_1-1} c^{\nu_2-1}\big[f(0,c) +f(a,0) -f(0,0)\big]=\frac{1}{\nu_1\nu_2} +\int_0^1\frac{da}{\nu_2a}(e^{-a\Si^2v_+}-1) +\int_0^1\frac{dc}{\nu_1c}(e^{-a\Si^2v_-}-1).
\end{equation}
Keeping in mind that
\begin{equation}
    \int_0^1\frac{da}{a}(e^{-za}-1)=-h(z),
\end{equation}
we arrive at
\begin{equation}
\begin{split}
    w^{(4)}_{s'_1s'_2}=&\,\frac{Z^2\al^2}{\bs\be'^2_e} w^{(1)}_{s'_1s'_2}\Big\{\int_0^1\frac{dadc}{ac}\Big[\frac{\exp\Big(-\frac{a\Si^2v_+ +c\Si^2v_- -2yu ac}{1-u^2 ac}\Big)}{1-u^2 ac} -e^{-a\Si^2v_+} -e^{-c\Si^2v_-} +1\Big]+\\
    &+\ln^2\frac{4\Si^2}{\la^2 e^\ga} -\big[h(\Si^2v_+)+h(\Si^2v_-)\big]\ln\frac{4\Si^2}{\la^2 e^\ga}\Big\}.
\end{split}
\end{equation}
Obviously, the integration over $|\spp'|$ can be performed by using formula \eqref{Gauss_int_rad} as in the case of the interference contribution. The cancelation of the infrared divergencies requires a careful separate study and will be given elsewhere. Notice that the standard arguments given, for example, in \cite{WeinbergB.12}, Chap. 13, rely on the assumption that the states of particles are described by plane waves and so these arguments have to be modified. Besides, in order to trace all the infrared divergencies up to the order $\al^2$ for the process at issue, one also needs to take into account the loop corrections of the respective order.


\end{document}